\documentclass[superscriptaddress,twocolumn,nofootinbib]{revtex4-2}
\usepackage{graphicx}
\usepackage{dcolumn}
\usepackage{bm}
\usepackage{mathrsfs}
\usepackage{multirow}
\usepackage{amsmath}
\usepackage{amsfonts}
\usepackage[makeroom]{cancel}

\usepackage[colorlinks,
linkcolor={red!60!black!},
anchorcolor={green!80!black!},
urlcolor={blue!80!black!},
citecolor={blue!50!black!}]{hyperref}
\usepackage[table]{xcolor}
\usepackage{tipa}
\usepackage{physics}
\usepackage{caption}
\usepackage{subcaption}
\usepackage{amssymb}
\usepackage{placeins}

\usepackage[linesnumbered]{algorithm2e}
\newenvironment{pseudocode}[1][htb]
  {
   \begin{algorithm}[#1]%
  }{\end{algorithm}}

  \def\ba{\begin{eqnarray}}
	\def\ea{\end{eqnarray}}

\def\q{\quad}

\begin{document}

\title{Causal structure and light-conical singularities in Lorentzian simplicial quantum gravity}

\author{Bianca Dittrich}
\email{bdittrich@pitp.ca}
\affiliation{Perimeter Institute, 31 Caroline Street North, Waterloo, ON, N2L 2Y5, Canada}

\author{José Padua-Argüelles}
\email{jpaduaarguelles@pitp.ca}
\affiliation{Perimeter Institute, 31 Caroline Street North, Waterloo, ON, N2L 2Y5, Canada}
\affiliation{Department of Physics and  Astronomy, University of Waterloo, 200 University Avenue West, Waterloo, ON, N2L 3G1, Canada}

\begin{abstract}

{\center \textbf{Abstract} }

The definition of a Lorentzian gravitational path integral requires specifying which causal structures are admitted off shell. Of particular importance are light-conical singularities, codimension-two irregularities of the light cone structure that render the gravitational action complex and can therefore lead to exponential enhancement or suppression of configurations. We investigate the prevalence and structure of such singularities in Lorentzian Regge gravity. We develop an efficient algorithm for randomly generating realizable Regge geometries satisfying the generalized triangle inequalities, and use it to sample the configuration space of four-dimensional triangulations. We find that light-conical singularities are generic: for bones shared by many four-simplices, configurations with irregular light cone structure overwhelmingly dominate, with trouser-type singularities generally entropically favored. Restrictions on the causal character of subsimplices, such as requiring all tetrahedra to be spacelike, can substantially alter these statistics. We further show that both yarmulke- and, in four dimensions, trouser-type singularities can occur on unbounded regions of configuration space associated with simple refinement moves. Beyond light-conical singularities, we analyze the causal structure of Lorentzian triangulations more generally, for example by introducing the notion of pairwise embeddable chrono-topologies and a set of discrete causality data. Our results demonstrate that causal irregularities are an intrinsic and abundant feature of the unrestricted Lorentzian Regge configuration space, sharpening the question of which causality conditions should be imposed in Regge gravity, spin-foam models, and ultimately the continuum gravitational path integral.

\end{abstract}

\maketitle


\section{Introduction}

The gravitational path integral has proven to be an extraordinarily useful tool for probing the quantum mechanics of gravitating systems. This is particularly true in the semiclassical regime, where saddle points dominate and geometric intuition remains a reliable guide. Saddle point evaluations, and in some cases the integration of fluctuations around them, have yielded deep insight into \emph{e.g.} black hole thermodynamics and quantum cosmology \cite{Gibbons:1976ue,Gibbons:1994cg,Ross:2005sc,Almheiri:2020cfm,Halliwell:1989myn,Lehners:2023yrj}. Yet, the path integral encodes quantum properties not in the stationary condition itself, but in the off-shell structure of its histories. Indeed, the relevance of a given saddle point can itself depend on such structure \cite{Marolf:2022ybi,Dittrich:2024awu,HalliwellContours}. Likewise, entropic properties of the space of off-shell configurations (such as their abundance or scarcity) may ultimately determine the large-scale, coarse-grained behavior of the system, substantially altering semiclassical expectations.

Thus, any path integral approach to quantum gravity needs to face, at some point, the question of what is the underlying off-shell structure. Understanding this structure, in turn, provides a window into deep quantum gravitational properties. These questions are particularly interesting when dealing with real-time, or Lorentzian approaches, where the difficult question of the fate of causality off-shell appears.

Discretizations of the gravitational path integral allow the exploration of such non-perturbative issues \cite{Loll:1998aj}. In this paper we will materialize them in the context of Lorentzian quantum Regge calculus, a non-perturbative lattice formulation of quantum gravity in which spacetime is approximated by a piecewise-flat simplicial manifold and the gravitational path integral is defined as an integral over possible geometries for the manifold.

Take for example the question of whether light cone structure degenerates off-shell —by degeneracy we mean, for example the possibility that a metric assigns two future and two past light cones to a point —\footnote{The authors like to use em dash, yet are humans.}see \emph{e.g.} FIGURE \ref{fig:trouserlike}. 

The real-time and discrete approach to quantum gravity of Causal Dynamical Triangulations (CDT) is defined by a configuration space obtained by gluing simplices with a fixed geometry following certain rules that ensure a well defined Wick rotation.\footnote{There are two versions of CDT: One with stricter rules, which can be interpreted as introducing an explicit slicing and imposing a proper time gauge on the triangulations \cite{Ambjorn:1998xu}. It can also be described as a dynamical triangulation with product structure \cite{Dittrich:2005sy}, where one of the two products is the triangulated line, representing time. This structure implies a regular light cone structure. A less strict version \cite{Jordan:2013awa,Jordan:2013iaa} imposes more directly a regular light cone structure (all spacetime points carry exactly one future-directed light cone and exactly one past-directed light cone) without an obvious slicing into hypersurfaces. Simulations indicate that these two versions are, in a statistical sense, in the same universality class.} These rules therefore exclude light cone degeneracies.

The situation is quite different in the Lorentzian approach to quantum gravity known as causal sets \cite{Bombelli:1987aa,Surya:2019ndm,Sorkin:2003bx,Dowker:2024fwa}. Here one encodes geometry into causal order and volume information. Spacetime is discretized into a set of so-called events and every event has by construction a well defined future and past. Beyond this, however, no further restrictions are imposed on the causal structure, and causal sets can approximate spacetimes with degenerate light cone structure \cite{Dowker:1999wu,Dowker:2002hm}.

The fact that different Lorentzian approaches make different choices already suggests that the status of such degeneracies is far from settled. Moreover, recent results indicate that spacetimes with light cone degeneracies of a certain type, which we will refer to as light-conical singularities (\emph{cf.} \S\ref{sec:lightconical_singularities}), may in fact be necessary for deriving thermodynamic quantities, such as entropies, from the Lorentzian path integral, both in the continuum \cite{Colin-Ellerin:2020mva,Marolf:2022ybi,Jacobson:2022jir} and in discrete (Regge) settings \cite{Dittrich:2024awu}. They also seem to be necessary for describing the emergence of no-boundary wavefunctions in Regge cosmology \cite{Dittrich:2021gww,Asante:2021phx}.

The mechanism underlying both observations is that light-conical singularities render the Einstein–Hilbert action complex \cite{Sorkin:1975ah,Sorkin:2019llw,Neiman:2013ap,Neiman:2013lxa}. This makes it possible to deform a real-time contour so that it passes through Euclidean saddles in a way that renders them semiclassically relevant \cite{Asante:2021phx,Marolf:2022ybi}. At the same time, however, the resulting path integral exponent can exponentially enhance certain histories, potentially leading to divergences.

Light-conical singularities therefore appear to be both physically useful and potentially problematic. Whether they should be included in the path integral, and under which conditions, remains a largely open question \cite{deBoer:2022zka,Buoninfante:2024yth}. This is perhaps unsurprising given the intrinsically non-perturbative nature of the problem. Yet, because the answer may profoundly affect macroscopic physics, it is also an imperative one. Unfortunately, current tools for exploring non-perturbative path integrals, particularly the interplay between action, measure, and entropic contributions, remain quite limited.

This is just one example of the physical consequences of the off-shell structure of the path integral, and of our incomplete understanding of it. Other examples include questions such as whether one should include histories with different unexpected causal features, histories that change spatial topology, or even histories that are non-manifold-like.

A first step to explore these issues is a general study of the configuration space of (Lorentzian) gravity to understand very generally the types of off-shell configurations one can have. 

The main aim of this paper is to begin such exploration with the discrete approach to quantum gravity of quantum Regge calculus. To do so, we first develop an algorithm to explore in a very general way, the Lorentzian Regge configuration space and then we will apply it to study light-conical singularities. Specifically we will address the key issues of how frequently they occur in configuration space.

The algorithm is designed as a computationally efficient tool for exploring the Regge configuration space. The Regge path integral naturally comes with a characteristic function restricting the integration domain to configurations satisfying the so-called generalized triangle inequalities. This suggests a straightforward sampling strategy: generate configurations at random and retain only those that satisfy the inequalities. In practice, however, a randomly generated configuration will only very rarely satisfy the generalized triangle inequalities,\footnote{For example, for simulations like the one in FIGURE \ref{fig:mixed_Delta_n} with $n=20$, where 20 4-simplices are glued cyclically around a common triangle, approximately 99\% of the samples do not satisfied the inequalities, for a large enough sample.} making this rejection-sampling procedure prohibitively inefficient for obtaining reliable statistics for large triangulations.

The algorithm we develop produces Regge configurations which satisfy the (Euclidean or Lorentzian) generalized triangle inequalities by construction. This works by applying a random sampling to generate embeddings of the simplices into flat spacetime,\footnote{Note that it is not the entire triangulation which is embedded into flat spacetime, but each single $d$-simplex.} with the constraint that they can be consistently glued together. 

We believe that the algorithm can be applied in many more contexts, beyond the application presented in this paper. 

In this paper we will also adress the question of whether light-conical singularities can appear with an unbounded integration domain. If we allow for an exponential enhancement of such configurations, an unbounded integration domain can lead to a divergent path integral, as previously mentioned.\footnote{This might however not always be the case: \cite{Dittrich:2024awu} presents a Regge micro-superspace model where one has a light-conical singularity with unbounded support, with an exponentially enhanced integration contour. Nevertheless the path integral can be made finite. This seems to depend however on the order of integration. The integral over the variable which controls the size of the $(d-2)$-simplices supporting the light-conical singularity (which in this case coincides with the size of the horizon) is performed last. The path integral over the other variables, which is performed first, leads to a destructive interference and ensures that the last integral has finite support.}
This investigation will be based on the construction of examples in order to see what kind of configurations can exist in principle. 

More generally, the question of which causal structures, and in particular which causal irregularities, can appear in Lorentzian Regge gravity remains largely unexplored \cite{Jordan:2013awa,Asante:2021phx}.\footnote{One exception is \cite{Asante:2025qbr}, which analyzes the causal structure around vertices in $(2+1)$-dimensional Regge triangulations.} To help address this gap, we go beyond light cone irregularities and study the causal structure of Lorentzian Regge triangulations more broadly. For example, we introduce the notion of \emph{pairwise embeddable chrono-topologies}, which roughly captures whether a triangulation admits a global time orientation, and illustrate it with various (counter-)examples. We also propose a data assignment that encodes the causal structure of a Lorentzian triangulation.

Having a deeper understanding of possible causal structures will help to decide what kind of causality constraints to impose on path integral for Regge gravity as well as related frameworks such as spin foams \cite{Perez:2012wv, Engle:2007wy,Freidel:2007py,  Asante:2020qpa, Asante:2021zzh, Livine:2024hhc}, and, via a continuum limit, for the path integral over continuum metrics. 

This paper is structured as follows. In Section \ref{sec:Regge_101} we review the basics of Euclidean and Lorentzian Regge calculus. Section \ref{sec:algorithm} introduces an algorithm for sampling the space of realizable geometrizations, \emph{i.e.} geometries on a triangulation with fixed connectivity that satisfy the generalized triangle inequalities. In Section \ref{sec:lightconical_singularities} we introduce the notion of light-conical singularities and present statistics on their occurrence in different set-ups. Section \ref{Sec:LargeBulk} provides explicit examples of light-conical singularities with unbounded integration domain, in particular for arbitrarily long bulk edge lengths. Such singularities are supported on codimension-two building blocks, \emph{i.e.} triangles in four dimensions, and in section \ref{Sec:Other} we briefly review other types of light cone irregularities, supported on edges and vertices. Section \ref{sec:chrono} addresses irregularities in the causal structure across codimension-one simplices, \emph{i.e.} the non-existence of a consistent time orientation on the triangulation. There we introduce the notion of pairwise embeddable chrono-topology and propose a data structure encoding such chrono-topologies. Finally, Section \ref{sec:discussion} contains our discussion and outlook.

\section{Regge geometries \label{sec:Regge_101}}

To facilitate the reading of the paper, we introduce in this section some basic background on Regge geometries.

For the purposes of this work, the fundamental building blocks of Regge geometries will be flat\footnote{One can also generalize to homogeneously curved simplices \cite{Bahr:2009qc,Bahr:2009qd}.} simplices. Morally, they are generalization of triangles and tetrahedra to any dimension. A $k$-dimensional simplex, or a $k$-simplex is defined as the convex hull of $k+1$ affinely independent\footnote{As a reminder, a set of points $\{p_0,\dots p_n\}\subset\mathbb R^{n}$ is said to be affinely independent if the vectors $p_1-p_0,\dots,p_n-p_0$ are \emph{linearly} independent.
 }
 points in $\mathbb R^k$. Equipping $\mathbb R^k$ with a flat Euclidean or Lorentzian metric induces a Euclidean or Lorentzian geometry, respectively, onto the simplex. This geometry\footnote{As we will see there are more data one might wish to include, \emph{e.g.} orientation and, for the Lorentzian case, time orientation.} can be reconstructed from the lengths associated to the edges of the simplex. Note that applying an isometric transformation will change the embedding of the simplex (\emph{i.e.} the points defining it), but not the geometry of the simplex. 

It is therefore often more convenient to understand simplices differently: as abstract objects (that is, not embedded in $\mathbb R^k$), characterized by a set of labeled vertices and geometric data, which we refer to as a \emph{geometrization}. 

Before commenting more on the concept of geometrization we need to introduce some additional notions. We note that any subset of vertices of a given simplex, defines, via its convex hull, a subsimplex. Its dimension is given by the number of vertices in the subset minus one. We will refer to the codimension-one subsimplices as \emph{faces}, and the codimension-two simplices as \emph{bones}.

The set of all (unordered) pairs of vertices of a $d$-simplex defines the set of its edges. It is customary to geometrize the simplex by assigning lengths to these $\binom{d+1}{2}$ edges.\footnote{Other choices can be found in \emph{e.g.} \cite{Barrett:1994nn, Bahr:2009qd, Asante:2018wqy,Dittrich:2023ava}.} From these lengths one can redefine a flat metric, see for instance \cite{Sorkin:1975jz,Asante:2021zzh}, and thus compute all possible geometric quantities for the simplex \cite{Borissova:2023izx}. However, not all edge-length assignments, that is geometrizations, guarantee that an embedding exists. If a geometrization can be realized via an embedding into flat space, we speak of a \emph{realizable} geometrization.

\begin{figure}
    \centering
    \includegraphics[width=0.5\textwidth]{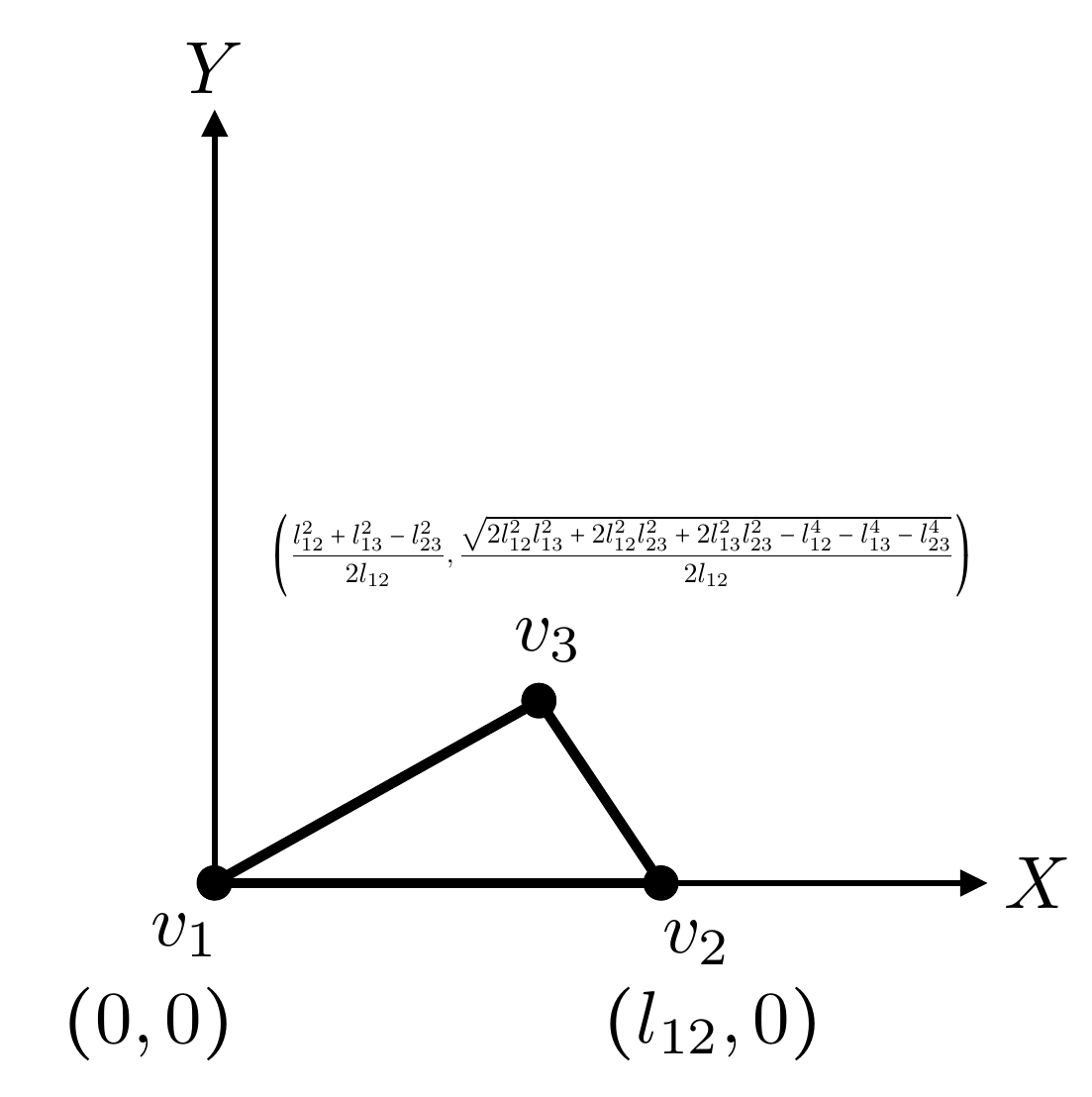}
    \caption{A choice of (`gauge' fixed) coordinates for an embedded triangle (in the Euclidean plan) with edge lengths $l_{ij}$. The embedding exist, \emph{i.e.} the coordinates are real, if and only if the triangle inequalities are satisfied.}
    \label{fig:triangle_embedding}
\end{figure}

The conditions of realizability can be written as a set of inequalities \cite{blumenthal1953theory,Tate:2011ct,Asante:2021zzh}, which generalize the triangle inequalities. Indeed, the Euclidean triangle inequalities, $l_{ij}< l_{ik}+l_{jk}$, determine whenever a geometrized triangle is realizable or not: As shown in FIGURE \ref{fig:triangle_embedding}, the coordinates of a Euclidean triangle with edge lengths $l_{12}$, $l_{13}$ and $l_{23}$ are sensibly defined (meaning real, and leading to an affinely independent vertex set), thus making the geometrization realizable, if and only if the Euclidean triangle inequalities are satisfied. 

Note that in FIG. \ref{fig:triangle_embedding} there is a kind of gauge that was chosen: namely $v_1$ is in the origin, the edge $\overline{v_1v_2}$ lies on the positive $X$-axis, and $v_3$ lies on the half-plane with $Y>0$. The associated gauge freedom is due to the aforementioned fact about isometric transformations leaving a simplex geometry invariant. With this gauge one can uniquely determine the coordinates of the vertices.

The procedure above can be generalized to any dimension, and it remains algebraically simple up to 4-simplices.  This suffices for the purposes of the algorithm described in the next section. We should point out, however, that when dealing with Lorentz signature, one needs to pay special attention to the signature of the subsimplices,\footnote{The signature of subsimplices can be read from their Cayley-Menger determinant, explained in the next paragraphs —see expression \eqref{eq:Cayley_Menger}.} when choosing the gauge fixing. So for example, if the triangle in a tetrahedron is timelike, the gauge needs to be such that the triangle is embedded in a timelike plane.


The \emph{Lorentzian} version of the triangle inequalities can be derived by computing a gauge fixed embedding analogous to the one in FIG.~\ref{fig:triangle_embedding}, and checking when the embedding results in real coordinates. The types of triangles and their corresponding inequalities are:

\begin{itemize}
    \item SSS: All edges are spacelike. Then there must be one edge larger than the sum of the other two. 
    \item SST: Two edges are spacelike and one is timelike. All geometrizations of this type are realizable.
    \item TTS: Two timelike edges and one spacelike edge. All geometrizations of this type are realizable.
    \item TTT: All edges are timelike. Here too, there must be one edge larger than the sum of the other two.
\end{itemize}

If one has at least one null edge, then any geometrization is realizable as long as the other two edges are different. 

In arbitrary dimension, realizability conditions for both Euclidean and Lorentzian simplices can be stated recursively. One requires that all proper subsimplices be realizable and, in addition, that the Cayley–Menger determinant,
\begin{equation}
    \frac{(-1)^{d+1}}{2^d (d!)^2}\det
    \begin{pmatrix}
    0 & 1 & 1 & 1 & \cdots & 1 \\
    1 & 0 & \ell_{12}^s & \ell_{13}^s & \cdots & \ell_{1(d+1)}^s \\
    1 & \ell_{12}^s & 0 & \ell_{23}^s & \cdots & \ell_{2(d+1)}^s \\
    1 & \ell_{13}^s & \ell_{23}^s & 0 & \cdots & \ell_{3(d+1)}^s \\
    \vdots & \vdots & \vdots & \vdots & \ddots & \vdots \\
    1 & \ell_{1(d+1)}^s & \ell_{2(d+1)}^s & \ell_{3(d+1)}^s & \cdots & 0
    \end{pmatrix}.
    \label{eq:Cayley_Menger}
\end{equation}
computed from the signed squared edge lengths $\ell_{ij}^s$, be positive in the Euclidean case and negative in the Lorentzian case. In the Lorentzian setting, one may also wish to allow null subsimplices, whose Cayley–Menger determinant vanishes.

For a realizable simplex, one can show (by constructing an embedding) that the Cayley-Menger determinant computes the signed volume squared of the simplex. Here and below we use ``volume'' in a loose way, to be adapted to the dimension of the object we are referring to; so for example `volumes' of triangles are to be understood as areas. Thus, the sign condition is reasonably asking the top-dimensional simplex to be spacelike in Euclidean signature and timelike in Lorentzian signature.

By using this characterization, one easily shows that 

\begin{itemize}
    \item A Euclidean simplex is realizable if and only if it has positive signed volume squared and all of its proper subsimplices do too.

    \item A Lorentzian simplex (without null subsimplices) is realizable if and only if its signed volume squared is negative and moreover, if a subsimplex has negative signed volume squared (\emph{i.e.}, it is timelike) then any (sub)simplex  containing it must also be timelike.
    
    If one allows null subsimplices, the condition for a Lorentzian $d$-simplex is that any (sub)simplex containing a null simplex must be null or timelike. The $d$-simplex itself has to be timelike.
\end{itemize}

As this new characterization might suggest, it is typically easier to work with signed squared geometric quantities, henceforth referred to  as squared length, squared volume, etc.

A $d$-dimensional Regge geometry consists of a collection of realizable $d$-simplices glued pairwise along shared faces, with the gluing condition requiring that the induced geometrization of each shared face agrees when computed from either of the adjacent simplices. It is this gluing that makes a Regge geometry non-trivial: Even though the simplices are flat themselves, they can be glued in a way that results in a curved triangulation.

\begin{figure}
    \centering
    \includegraphics[width=0.4\textwidth]{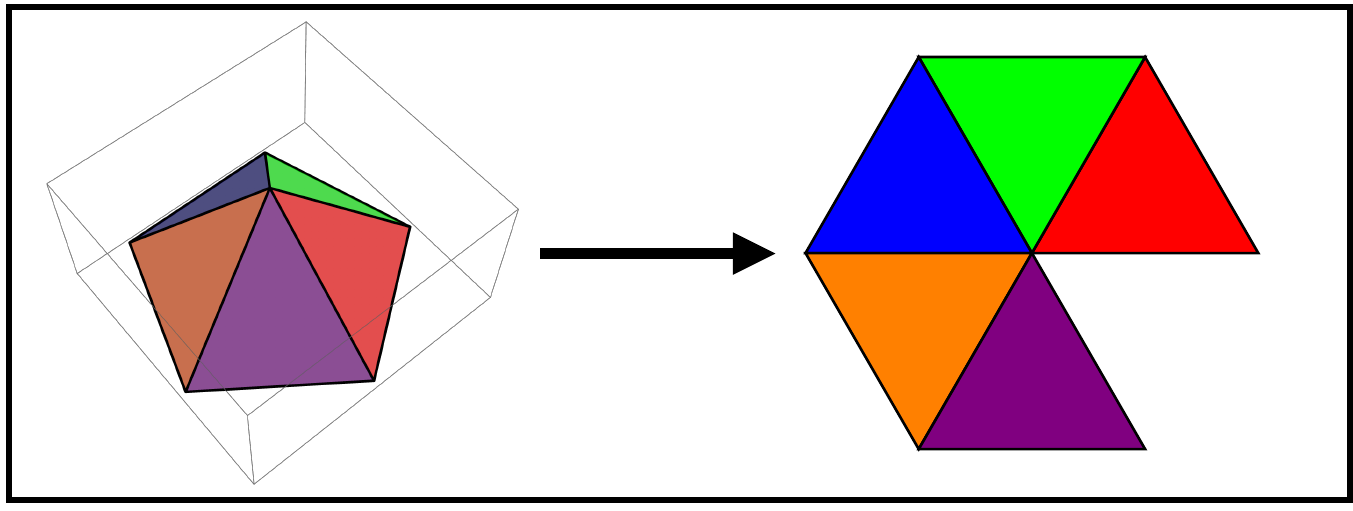}
    \caption{The deficit angle around a bone (here, the bone is represented by the center point of the triangulation on the left) determines when the bone circuit, \emph{i.e.} the triangles around the bone, can be embedded into flat spacetime without breaking the circuit. In the picture, the deficit angle is non-zero, and thus the circuit is broken in the flat spacetime embedding on the right. }
    \label{fig:deficit_angle}
\end{figure}

More specifically, the geometric data induce for the bones of a triangulation a \emph{deficit angle}, which when non-zero implies that the complex given by all the simplices sharing the bone cannot be embedded in flat spacetime. A visualization of the concept behind deficit angles is shown in FIG.~\ref{fig:deficit_angle} for a two-dimensional triangulation. The deficit angle of a bone $b$ is defined as \cite{Regge:1961px, Sorkin:2019llw}
\begin{equation}
    \epsilon=\theta_\text{flat}-\sum_{\sigma\supset b}\theta_{\sigma\supset b},
\end{equation}
where $\theta_\text{flat}$ is the angle covered by the whole flat plane in the corresponding signature —so \emph{e.g.}  $\theta_\text{flat}=2\pi$ in Euclidean signature, and  $\theta_\text{flat}=-\imath 2\pi$ in Lorentzian.\footnote{\label{fnote:sign_ambiguity} There is a sign ambiguity in the overall sign of $\theta_\text{flat}$, which is of particular relevance in Lorentz signature. This ambiguity leads to the same Regge action for lightcone regular configurations (see \S\ref{sec:lightconical_singularities}), but can lead to opposite signs for imaginary terms for configurations with light-conical singularities \cite{Asante:2021phx}. To simplify the discussion we adopted one choice of sign for $\theta_\text{flat}$. We should keep in mind however, that using the complex Regge action, the different sign choices are only relevant along branch cuts, where these signify a choice of side along the cut. (See also footnote \ref{fnote:Halliwell-Hartle_ambiguity}.)} The sum is over all simplices $\sigma$ sharing the bone, and $\theta_{\sigma\supset b}$ is the \emph{dihedral angle} of $b$ in $\sigma$: \emph{i.e.} the \emph{inner} angle between the hyperplanes defined by the faces of $\sigma$ that meet at $b$ —in any embedding. 

The dihedral angles $\theta_{\sigma\supset b}$ can be computed by constructing an embedding of the simplex in question, and projecting onto the plane orthogonal to the bone.\footnote{Note that bones are codimension-two, so their orthogonal complement is always two-dimensional.} In this way, the simplex gets projected into a triangle with one vertex resulting from the projection of the bone, and its neighboring edges resulting from the projection of the faces which share the bone. The (internal) dihedral angle between these faces is equal to the (internal) angle of the projection triangle at the vertex that corresponds to the projected bone.

Let us note that Lorentzian angles (which appear for spacelike bones), are in general complex-valued. See \emph{e.g.} \cite{Sorkin:2019llw, Asante:2021phx} for detailed derivations.  Within the conventions we use here the angles can have an imaginary part which is a multiple of $-\pi/2$, where the multiple counts the number of light ray crossings included in the angle. Because simplices are always convex, Lorentzian dihedral angles can include either zero or two light ray crossings, if the dihedral angle is between two faces of the same signature, or one light ray crossing if the dihedral angle is between a timelike and a spacelike face.

The deficit angle provides a notion of curvature-per-area, which, as Regge showed in \cite{Regge:1961px}, can be used to define a discrete gravitational action. (See \cite{Sorkin:1975ah} for the first construction of the Lorentzian Regge action.) The associated dynamics is consistent with Einstein's gravity in the sense that in the limit of an infinitely refined triangulation, it reproduces the continuum dynamics encoded in the Einstein-Hilbert action —for more detailed discussions, see \cite{Williams:1991pj,Misner1973,Rocek:1981ama,Rocek:1982tj,barrett1988convergence,Barrett:1988wd,Feinberg:1984he,Sorkin:1975ah,Friedberg:1984ma,osti_4042676,Gentle_1998,Brewin:2000zh, Bahr:2009qc, Bahr:2010cq}. The discrete action is known as Regge action $S_\text{R}$. In absence of a cosmological constant and boundaries, and in units in which $8\pi G=1$, it is given by \cite{Regge:1961px,Sorkin:1975ah,Sorkin:2019llw, Asante:2021phx}
\begin{equation}\label{ReggeAction}
    S_\text{R}=\sum_b\text{Area}(b)\epsilon_b.
\end{equation}
where $\text{Area}(b)$ is the (positive) area of the bone.\footnote{The area for spacelike bones is defined as square root of the signed squared volume of the bone (which can be computed via the Cayley-Menger determinant), and for a timelike bone as the square root of minus the signed squared volume. The area for a null bone is vanishing, null bones do therefore not contribute to the Regge action.} All quantities are to be understood as functions of the squared edge lengths.\footnote{Other choices are possible, but may lead to a different dynamics or require additional constraint terms, see \emph{e.g.} \cite{Barrett:1994nn, Bahr:2009qd, Asante:2018wqy}.}

Motivated by the fact that in the refinement limit, Regge dynamics reproduces Einstein's, one can aim to define the quantum gravitational path integral by taking a refinement limit of the Regge  path integral
\begin{equation}
    Z_\triangle=\int_{\Omega_\triangle}\mathrm d^N l_i^s\mu(l^s_i) e^{\imath S_\text R} .
\end{equation}
Here the integral over edge lengths squared $l^s_i$ of the triangulation $\triangle$, provides a discrete analog of metric variables, and $\mu$ is a gravitational measure. For the integration domain  $\Omega_\triangle$ one often considers  all possible geometrizations (of a triangulation with fixed connectivity), restricted only by the triangle inequalities.\footnote{Alternatively one might complement $\mu$ with a characteristic function of the space that satisfies the triangle inequalities and integrate over all geometrizations.} As we have hinted at, and will keep discussing throughout the text, for Lorentzian signature one could also consider to impose certain regularity conditions for the causal structure. This is however a very open question, as we have a poor understanding of how such regularity conditions would impact the path integral predictions. 

To make progress, part of this work will explore typicality statements for certain types of causal irregular configurations. We will explore these for the configuration space given by all realizable geometrizations , that is, we will only impose the generalized triangle inequalities.

\section{Algorithm to produce realizable geometrizations \label{sec:algorithm}}

We now introduce a useful tool for exploring the configuration space of Regge histories: an algorithm that randomly generates realizable geometrizations for the Regge path integral. For generic triangulations, this enables much more efficient simulations compared to randomly assigning lengths and then filtering for those satisfying the generalized triangle inequalities. This will be essential to allow for the simulations in §\ref{sssec:Delta_n}.

Although the core ideas of the algorithm extend to triangulations in arbitrary dimension $d$, we focus on the physically relevant case $d=4$, which is also the dimension used in the simulations of \S\ref{ssec:simulations}. Likewise, we restrict our attention to Lorentzian signature. For simplicity, we exclude configurations containing null (sub)simplices, which form a measure-zero subset of the configuration space and also require additional care —\emph{cf.} \ref{sec:chrono}. 

The main step in the algorithm is iterative: We assume that we have given embedding data for a set of simplices and construct explicit embeddings for adjacent simplices. Initially, the embeddings would typically be induced by boundary conditions or by prescribed geometric data for the bulk of the triangulation but could also be generated randomly —\emph{e.g.}  if the triangulation has no boundary and bulk constraints of interest.

In applying this iterative step the algorithm picks a bone and goes through the 4-simplices in the bone's circuit one-by-one. After having constructed a geometrization for the bone circuit the algorithm picks another bone and proceeds with the geometrization of its circuit. 

\begin{figure}[h!]
    \centering
    \begin{subfigure}[c]{0.65\columnwidth}
        \centering
        \includegraphics[width=\linewidth]{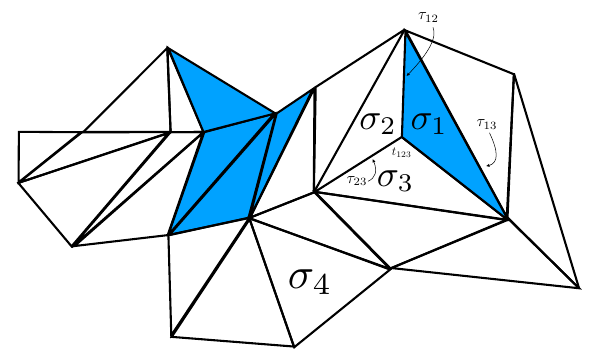}
        \caption{Constructing embeddings for the 4-simplices sharing the triangle $t_{123}$, \emph{i.e.} the bone circuit of $t_{123}$, at a stage where the only geometrized simplex in the circuit of this triangle is $\sigma_1$. The algorithm proceeds to construct a geometrization for $\sigma_2$ as described in the main text.}
        \label{fig:random_point_step}
    \end{subfigure}
    \hfill
    \begin{subfigure}[c]{0.65\columnwidth}
        \centering
        \includegraphics[width=\linewidth]{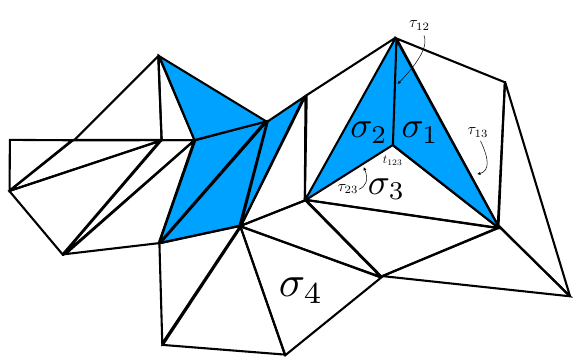}
        \caption{Constructing embeddings for the 4-simplices in the bone circuit of $t_{123}$ at a stage where $\sigma_1$ and $\sigma_2$ have already been geometrized. The algorithm proceeds by creating an embedding for $\sigma_3$ using the dihedral angle at $t_{123}$.}
        \label{fig:consistent_embedding_step}   
    \end{subfigure}
    \caption{Illustration of some steps and aspects of the algorithm used to construct realizable geometrizations for a given triangulation. The triangles represent 4-simplices $\sigma_i$ and correspondingly, the edges and vertices depict tetrahedra $\tau_{ij}$ and triangles $t_{ijk}$, respectively. Colored triangles denote simplices whose geometry has already been specified. When exploring the triangle $t_{123}$, the algorithm goes around its circuit to sequentially construct geometrizations for $\sigma_2$ and $\sigma_3$, starting from the already existing geometrization of $\sigma_1$. Note that although $\sigma_4$ has not been fully geometrized, a triangle of it has —represented in the figure by its vertex shared with the colored triangles.}
    \label{fig:steps}
\end{figure}

To give a basic example for the bone circuit geometrization, suppose that in some triangulation, a realizable assignment has  already been given to a simplex $\sigma_1$, but not yet to a neighboring simplex $\sigma_2$, as illustrated in FIG.~\ref{fig:random_point_step}. Because the simplices are adjacent, some geometry has already been assigned to $\sigma_2$, namely the one inherited from the shared tetrahedron $\tau_{12}$. Assume that this is the only geometry given to $\sigma_2$ so far. One can then construct an embedding (and therefore a geometrization) of $\sigma_2$ by implementing the following steps:
\begin{enumerate}
    \item Embed $\tau_{12}$ in four dimensional spacetime, which must be possible since $\sigma_1$ carries a realizable geometrization. This provides coordinates for four vertices of $\sigma_2$.
    \item Generate the coordinates of the fifth vertex randomly.\footnote{We will comment more on random samplings at the end of this section.}
\end{enumerate}

This example shows how the geometrization of the triangulation can be grown out of already existing geometric data. The process is continued by focussing on the next simplex in the bone circuit. Assume we are in the situation of FIG.~\ref{fig:consistent_embedding_step}, and thus such simplex is $\sigma_3$, which shares tetrahedra $\tau_{13}$ and $\tau_{23}$ with $\sigma_1$ and $\sigma_2$, respectively. The only geometric datum we have not specified yet for $\sigma_3$ is the length of the edge  opposite to the triangle $t_{123}$, which is shared by $\tau_{13}$ and $\tau_{23}$. (We assume that this edge has not yet been assigned a length by \emph{e.g.} fixed bulk data.) Then, the algorithm generates a (squared) length for it, such that the generalized triangle inequalities are satisfied. This proceeds in three stages:

\begin{enumerate}
\item
First, compute the lengths of the edges obtained by projecting out $t_{123}$ from $\tau_{13}$ and $\tau_{23}$ —\emph{e.g.} by embedding each tetrahedron in four-dimensional spacetime. Any embedding of the 4-simplex $\sigma_3$ would project to a triangle $\Pi\sigma_3$, whose two edges meeting at the projection of $t_{123}$ have precisely the lengths computed above.
\item
As we will show below, to determine a consistent length for the edge opposite to the bone, it suffices to provide a realizable geometrization for the triangle $\Pi \sigma_3$ resulting from the projection above. As we know already two of its edge lengths, we only need to reconstruct the third's length. We do so by randomly selecting a length
 in the space consistent with the relevant triangle inequalities. Note that we know the signature of the triangle $\Pi \sigma_3$: it has to be spacelike for a timelike bone and timelike for a spacelike bone.
\item 
From the geometry of the triangle $\Pi \sigma_3$ we can determine the dihedral angle $\theta_{123}$ of $\sigma_3$ at the triangle $t_{123}$. Given the lengths of all the other edges of $\sigma_3$ and the dihedral angle, we can compute the length of the remaining edge in $\sigma _3$. 
\end{enumerate}

To show the statement made in point 2., we provide a construction for the simplex $\sigma_3$. We start by embedding the complex given by the two tetrahedra $\tau_{12}$ and $\tau_{23}$ glued at $t_{123}$ into 4D Minkowski space. To obtain the correct dihedral angle $\theta_{123}$, we start with an embedding with a vanishing or an orthogonal dihedral angle, and then apply a rotation or boost to reach the actual value of  $\theta_{123}$. Here we need to distinguish between the cases where (a) the triangle $t_{123}$ is timelike or (b) the  triangle $t_{123}$ is spacelike. (We remind the reader that we excluded null subsimplices.)
\begin{itemize}
\item[(a)] If $t_{123}$ is timelike, the tetrahedra $\tau_{12}$ and $\tau_{23}$ are also timelike and we can embed them into a timelike 3D hyperplane, so that the dihedral angle is $\pi$. We then apply a rotation around $t_{123}$ to obtain an embedding with the correct value.
\item[(b1)] Assume that $t_{123}$ is spacelike and one of the tetrahedra is also spacelike but the other is timelike. Then, as the triangle $\Pi \sigma_3$ is convex and the dihedral angle $\theta_{123} $ is between a spacelike and a timelike edge, it includes one and only one light ray crossing and thus has an imaginary part $-\pi/2$. We embed the two tetrahedra $\tau_{12}$ and $\tau_{23}$ so that the dihedral angle between them is $-\imath \pi/2$. We then apply a boost (with a positive or negative boost parameter) to realize the actual value of $\theta_{123}$.

\item[(b2)]  If $t_{123}$ is spacelike and both  tetrahedra are spacelike, $\theta_{123}$ is a Lorentzian angle with imaginary part equal to $0$ or equal to $-\pi$. This corresponds to a thin Lorentzian angle (not containing a light cone) or a thick Lorentzian angle (containing one light cone) respectively. We embed $\tau_{12}$ and $\tau_{23}$ such that the dihedral angle between the tetrahedra is equal to the imaginary part of $\theta_{123}$. We then apply a boost to obtain an embedding with dihedral angle equal to $\theta_{123}$.
\end{itemize}

This process of constructing an embedding after consistently geometrizing two faces of a simplex meeting at a bone, which we will refer to as a \emph{dihedral construction}, generalizes to any dimension, because projecting out a bone always produces a triangle, regardless of the dimension. This makes dihedral constructions particularly useful for the algorithm: They are the main tool for building embeddings from partial geometrizations —both within and beyond the examples provided above.

Using these geometrization processes, the algorithm successfully constructs full histories for the Regge path integral by applying them to every bone circuit in the triangulation. This is the main idea behind the algorithm. We now turn to explain it in full detail.

First, we begin by defining some important terms we will use:
\begin{itemize}
    \item A \textbf{fixed (sub)simplex} is one whose geometry is fixed \emph{a priori}.
    \item An \textbf{embedded (sub)simplex} is one for which an embedding has been constructed.
    \item The \textbf{bone circuit} of a bone $b$ is the set of 4-simplices $\sigma_i(b)$ meeting at $b$, equipped with a cyclic ordering, $(\sigma_1(b),\dots,\sigma_{n_b}(b)$, so that $\sigma_i(b)$ and $\sigma_{i+1}(b)$ share a tetrahedron that contains $b$, and similarly for $\sigma_{n_b}(b)$ and $\sigma_1(b)$, and we identify  $\sigma_0$  with $\sigma_{n_b}$. In the context of working within a circuit, the tetrahedron that $\sigma_i(b)$ shares with $\sigma_{i-1}(b)$ ($\sigma_{i+1}(b)$) will be referred to as the \textbf{previous (next) tetrahedron}. These are always the two tetrahedra in $\sigma_i(b)$ that meet at $b$. 
    \item The main routine of the algorithm consist of visiting, or \textbf{exploring}, different bones and at each, construct embeddings following its circuit. A bone will be said to be \textbf{explorable} if it has not been explored, but lies in some embedded simplex. Note that by definition, every explorable bone $b$ has at least one simplex in its circuit that has already been embedded. Any such simplex will be used as the first of the circuit. Which embedded simplex is chosen is a random choice.
    \item Finally, the \textbf{star} of an edge is given by all the simplices sharing it.
\end{itemize}
\emph{Initialization of the algorithm:} The output of the algorithm is the list \texttt{embeddings}, consisting of  ordered pairs (simplex label, simplex embedding). The list is initialized by taking into account the fixed subsimplices. Geometry that is fixed \emph{a priori} will be assumed to be realizable. Fixed 4-simplices can thus be embedded by assumption and one only needs to construct these embeddings and add them to the list. Now, typically one is only fixing boundary geometry, which makes the situation more subtle as this usually boils down to defining some fixed tetrahedra. This information will be incorporated into \texttt{embeddings} by imagining that the tetrahedron is part of a virtual boundary simplex defined by the vertices of the tetrahedron and one extra virtual vertex. This virtual simplex can be embedded as was done with $\sigma_2$ in the example above, and this information will be the one added to \texttt{embeddings}. This \emph{extended} simplicial complex structure will play an important role when exploring boundary bones. 

Now,  one might also be dealing with a case in which no geometry is fixed \emph{a priori}, for example, if there are no bulk constraints and the triangulation has no boundary. In such a case, a simplex will be randomly selected and assigned a random embedding, \emph{i.e.} we generate five random points in flat spacetime. Then, this will be the first entry in \texttt{embeddings}.

\emph{Main body of the algorithm:} The main body of the algorithm is presented in the pseudocode \ref{pscode:coarse}.
\begin{pseudocode}[ht]
  \SetKwProg{Catch}{catch}{:}{}
  \SetKwProg{Throw}{throw}{:}{}
  \SetKwComment{Comment}{(*}{*)}
  \SetKwFunction{AttemptEmbedding}{AttemptEmbedding}
  \SetKwFunction{ConstructEmbedding}{ConstructEmbedding}
  
  \KwData{(1) A finite set of 5-element sets of vertex labels, each representing a 4-simplex.\\
  \hspace{3.2em}(2) Partial geometric data.}
  \KwResult{Consistent 4D embeddings for all simplices.}  
  \While{\text{embedded simplices$\neq$extended simplices}}{
    $b\gets$\text{RandomChoice(\emph{explorable bones})};\\
        \For{$i = 2$ \KwTo length of $b$'s circuit}{
            \If{$\sigma_i(b)$ has already been embedded}{
                Skip to next step in the for loop \textbf{3}
            }
            \ElseIf{Embedding of $\sigma_i(b)$ is unconstructible}{
                Erase existing embeddings associated to the inconsistent data;\\
                Update lists
            }
            \ElseIf{Embedding of $\sigma_i(b)$ should be postponed}{
                Go back to step \textbf{2}
            }
            \Else{
                Construct embedding of $\sigma_i(b)$ using its partial geometrization;\\
                Update lists
            }
            $i\gets i+1$\\
        }
        Update lists
  }
  \caption{Main structure of the algorithm used to construct realizable histories for the Regge path integral.}
  \label{pscode:coarse}
\end{pseudocode}
To explain the details of Pseudocode~\ref{pscode:coarse}, we will refer to the N-th line in the code as \ref{pscode:coarse}:\textbf{N}.  First, we can see that the \textbf{while} loop in \ref{pscode:coarse}:\textbf{1} ensures that the routine runs until all simplices have been embedded. Within each iteration of such loop, a particular bone will be explored, thanks to \ref{pscode:coarse}:\textbf{2}. Finally, while exploring a particular bone, the simplices of its circuit will be visited through the \textbf{for} loop in \ref{pscode:coarse}:\textbf{3}, and their embeddings will be constructed, if possible or needed. 

Modulo the initialization case discussed above, the strategy is to construct embeddings for 4-simplices $\sigma$ from pre-existing geometric data. This relies on the specific ordering of simplices within each bone circuit, as defined earlier: the ordering guarantees that in the \textbf{for} loop in \ref{pscode:coarse}:\textbf{3}, the embedding of $\sigma_{i-1}$ is already available when visiting simplex $\sigma_i(b)$.

It might be that some given $\sigma_i(b)$ has already been embedded —\emph{e.g.}  when some other bone was explored. That explains \ref{pscode:coarse}:\textbf{4}-\textbf{6}.

Previously explored bones, or fixed geometric data, may also determine parts of the geometry of $\sigma_i(b)$ beyond what is shared with $\sigma_{i-1}(b)$. If this geometry is inconsistent, in the sense that it might be related to an unrealizable (sub)simplex, corrective action is needed. This occurs because independent embeddings may assign conflicting lengths to shared subsimplices that have not yet been geometrically reconciled. In such cases, elements must be removed from \texttt{embeddings} to eliminate the conflicting length assignments. This is what \ref{pscode:coarse}:\textbf{7-9} takes care of. To avoid deleting too much progress, this can be done minimally: elements in \texttt{embeddings} associated with stars of edges of the problematic (sub)simplex are removed one edge at a time. Note that the update process not only affects \texttt{embeddings}, but also (potentially) other lists  —for example, it might erase the geometry of an edge in a previously explored bone, which makes this bone unexplored again. In this case it must be marked as such.

Whenever the edge opposite to $b$ in $\sigma_i(b)$  has been assigned a geometry, but the geometry of the next tetrahedron has not yet been fixed, a dihedral construction cannot be performed. In such cases, bone exploration is postponed (\ref{pscode:coarse}:\textbf{11–13}).

If the routine reaches \ref{pscode:coarse}:\textbf{14}, an embedding for $\sigma_i$ is constructed. Having discussed the preceding cases, it remains only to consider this one. For the remainder of this explanation of Pseudocode~\ref{pscode:coarse}, we therefore assume that the routine reaches \ref{pscode:coarse}:\textbf{14} and examine the possible scenarios that follow.

We begin with the situation in which the edge opposite $b$ in $\sigma_i(b)$ has already been geometrized. If that is the case and the routine reaches \ref{pscode:coarse}:\textbf{14}, then the next tetrahedron must also have been geometrized, as test \ref{pscode:coarse}:\textbf{11} failed. Then the entire simplex $\sigma_i(b)$ is geometrized —recall that the previous tetrahedron belongs to an already embedded simplex by construction. Moreover, this data is realizable, since \ref{pscode:coarse}:\textbf{8} was \emph{not} triggered. An embedding can therefore be constructed, and the algorithm proceeds accordingly —this can be done, \emph{e.g.} , by proceeding analogously to a dihedral construction, except that instead of generating the dihedral angle, one computes it from the geometric data of the simplex.

\begin{figure}
    \centering
    \includegraphics[width=0.5\textwidth]{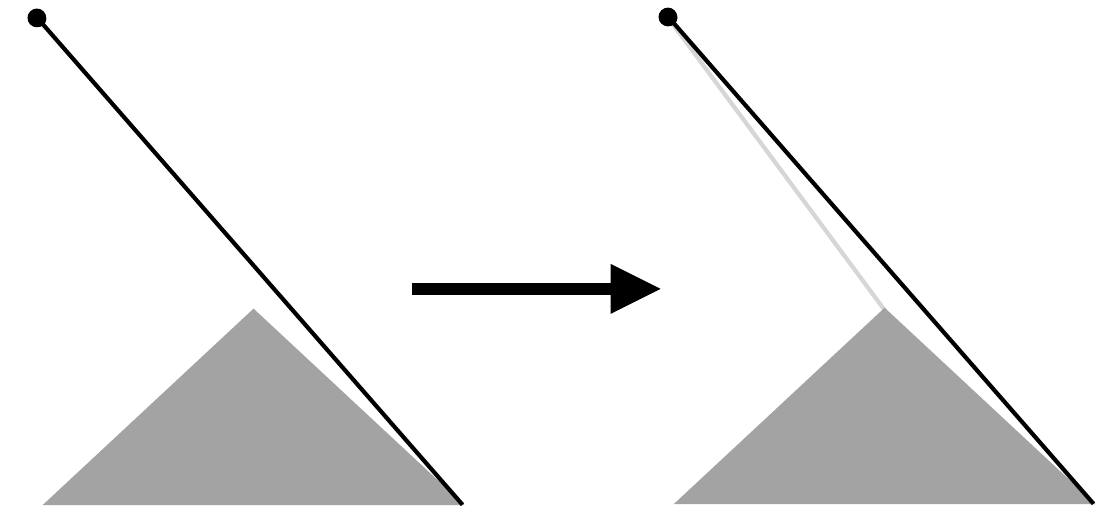}
    \caption{If only four edges of a tetrahedron have been geometrized, then a full triangle and an edge have been geometrized. If the triangle's geometric data is realizable, one can embed it. One of the vertices of the remaining edge must be one of the triangle's vertices in any embedding of the full tetrahedron. Thus one has a partial embedding of the tetrahedron, as shown in the left side. The embedding is partial, because the position of the highlighted vertex is not fixed yet, as it would be captured by the lengths of the edges in the tetrahedron that have not been geometrized. The algorithm proceeds by first setting the geometry of one of these edges in a way such that the new triangle that is formed is realizable. Later, the algorithm geometrizes the last edge using a 3D dihedral construction.}
    \label{fig:only_four}
\end{figure}

Now to the scenarios in which the opposite edge has not been geometrized. These are classified by the number of edges in the next tetrahedron whose geometry has been fixed. Since the three edges of $b$ are always geometrized (by construction, because $b$ is explorable), the possibilities are:
\begin{itemize}
    \item Only \emph{three} edges have been geometrized: Only the edges of the bone $b$ are known. In this case, the algorithm proceeds as it did for the example of $\sigma_2$ discussed earlier.
    \item Only \emph{four} edges have been geometrized: This case is illustrated in FIG.~\ref{fig:only_four}. The algorithm deals with this case by geometrizing the edge $e_{24}$, so that the triangle $\triangle_{124}$ is realizable. Then, we can continue by considering the case in which
    \item only \emph{five} edges have been geometrized: All edges of the tetrahedron are fixed except one. The algorithm does a 3D dihedral construction. In this way the two tetrahedra meeting at $b$ have been geometrized, so finally we perform a dihedral construction —as we have a situation equivalent to the one in the example of $\sigma_3$. We can then move to the case in which
    
    \item  all \emph{six} edges have been geometrized: The entire next tetrahedron is already geometrized. In this case, the algorithm does a 4D dihedral construction, —as was done with the example of $\sigma_3$.
\end{itemize}

This covers how the algorithm works. Note however, that for practical implementations, some of the steps above can be mixed for efficiency. For example, instead of running all of the required realizability tests in \ref{pscode:coarse}:\textbf{7}, it is more optimal to adequately mix these tests with the ones relevant for \ref{pscode:coarse}:\textbf{15}. A concrete implementation of the algorithm, incorporating some of these kind of optimizations, is included in the Mathematica package described in \cite{PaduaArguellesQRC:2026}.

~\\

The algorithm above is intended to explore the space of Regge geometries in the most general way possible. Meaning that it is meant to generate all possible length assignments of a triangulation, consistent with triangle inequalities,\footnote{Not all simplicial approaches to quantum gravity impose the full set of general triangle inequalities. For example, in spinfoams, some subset of the triangle inequalities is only imposed `weakly', \emph{i.e.} configurations that violate them may not give transition amplitudes that are vanishing, but are exponentially suppressed instead \cite{Ponzano:1968,Barrett:1993db,Barrett:2008wh}. This can have important consequences for cosmological applications as well as coarse graining procedures \cite{Dittrich:2023ava}. \label{fnote:spinfoam_realizability}} regardless of global or causal considerations. For example, we do not impose the geometry of the triangulation to have a time foliation —as done \emph{e.g.}  CDT or even in the Quantum Regge Calculus work of \cite{Tate:2011ct}. This allows to use the algorithm as an exploring tool of the most general configuration space for the Regge path integral. 

~\\
\emph{Remarks on the generated statistical distribution:}

The statistical distribution generated by the algorithm above is determined by three kinds of samplings: $(1)$ Firstly, bones or simplices are randomly picked throughout the algorithm. $(2)$ Some steps of the algorithm involve the sampling of vertex coordinates in $\mathbb{R}^d$, and $(3)$ the dihedral construction involves the sampling of a length squared —of an edge in the triangle resulting from projecting out a bone from a 4-simplex. For practical reasons, the two latter random processes must be defined on compact sampling spaces. For example, for $(2)$, one possibility, which we adopt below, is the uniform measure on a box $\left[-L,L\right]^d$ centered at the barycenter of the (sub)simplex under consideration. For $(3)$ we have two cases: either the triangle inequalities for the edge give a finite interval or they do not. If they do, we choose such interval. If they do not, we again introduce a cutoff scale $\tilde{L}$ to define a finite interval. In any case, once we have a finite interval to sample from, we do so using the uniform distribution.

Thus $L$ (and $\tilde L$) will provide a scale characterizing the resulting probability distribution. In the simulations below we will sample at different scales $L, \tilde L$ in order to be able to make statements on the limit distribution $L,\tilde L \rightarrow \infty$, which probes all realizable configurations.  However, setting \emph{e.g.} $\tilde L=L$, our results for finite $L$ are still meaningful for systems where there is a distinguished scale — provided for example by boundary conditions.

We note that because of $(1)$, that is the random selection of bones or other simplices in the sampling procedure, we are working with a probability distribution of measures. That is due to the fact, that different runs may geometrize the triangulation in different ways, \emph{e.g.} the dihedral construction will be applied to different edges. We thus avoid singling out the edges to which the dihedral construction is applied.

This choice of probability distribution allows for a sufficiently fast algorithm. Let us comment on other possible choices and issues to be addressed in future work.

One other reasonable choice for a probability distribution would be to use the Lebesgue measure on the space of edge lengths (or squared edge lengths). To this end one would to have to implement the Jacobians for the vertex sampling (this Jacobian involves the volume of the sampled simplex if one uses squared edge lengths) and the dihedral construction (this Jacobian is trivial if one uses squared edge lengths), which are derived in Appendix \ref{App:Jac}.  However, to determine the normalization one needs to implement all the generalized triangle inequalities, which can lead to very complicated boundaries of the allowed  configuration space —which in Lorentzian signature can be also disconnected. It is exactly this issue, which we wanted to avoid with our sampling procedure. 

Another possibility would be to use the Regge gravitational measure, which is also typically expressed in terms of (squared) edge lengths. However, its definition in 4D is still an open issue —see \cite{Loll:1998aj,Hamber:2009zz,Dittrich:2011vz, Dittrich:2014rha, Borissova:2023izx} for discussions. One question, which could influence the resulting distribution in an essential way, is whether this measure is singular on configurations which involve null subsimplices.

Finally, let us comment on the issue of diffeomorphism symmetry and possibly diffeomorphism equivalent configurations. In the authors' opinion there is no generally agreed upon definition of diffeomorphism equivalence for Regge configurations on the purely kinematical level \cite{Loll:1998aj, Dittrich:2008pw}. On the dynamical level, one can however determine whether solutions correspond to isolated extrema of the Regge action (in which case we do not have gauge, and in particular diffeomorphism symmetry for this solution) or correspond to a continuously connected $n$-parameter set of extrema (in which case we would speak of an $n$-dimensional gauge symmetry) \cite{Dittrich:2008pw}. For the 4D Regge action (without cosmological constant) flat solutions do feature such a gauge symmetry with $4\times V$ gauge parameters where $V$ is the number of bulk vertices.\footnote{In cases without boundary or a degenerate boundary one has to subtract the dimension of the global group of isometries from $4 \times V$ to obtain the number of gauge parameters. The isometries keep all the edge lengths the same but change the embedding of the vertices and thus the embedding of the entire triangulation.} In this case the various solutions can be understood to arise from the different placements of the $V$ vertices in flat space.
Solutions with curvature (more specifically without any bulk vertices inside a flat region) correspond generically to isolated extrema \cite{Bahr:2009ku} and thus do not feature gauge symmetries. 

Flat configurations occur with measure zero within our sampling procedure, we will therefore ignore here the issue of diffeomorphism invariance. 

There are however global isometries, that is rigid transformations, which do not change the lengths of any edge but do change the embedding of the vertices. This is taken into account by normalization factors when utilizing the algorithm to compute expectation values.

Having presented our algorithmic tool to explore a very general space of configurations for the Regge path integral, let us use it to explore some causal aspects of the Lorentzian theory. In particular, we will explore how generic it is for histories to have a `regular light cone structure' —in a specific sense defined below.

\section{Light-conical singularities \label{sec:lightconical_singularities}}

The two minimal constraints used to define the Regge configuration space so far are generalized triangle inequalities and matching conditions, \emph{i.e.} that two neighbouring simplices induce the same geometry on the subsimplex they share.\footnote{This is yet another feature (\emph{cf.} footnote \ref{fnote:spinfoam_realizability}) that can be relaxed in (four-dimensional) spinfoams \cite{Dittrich:2008ar,Dittrich:2010ey,Dittrich:2008va}.} The conditions are quasi-local as the generalized triangle inequalities apply independently to each 4-simplex and the matching conditions apply to pairs of neighbouring 4-simplices. This lack of global conditions can lead to simplicial geometries that seem exotic from a continuum and classical point of view. This can result in configurations with irregular light cone structure.

\begin{figure}[h!]
    \centering
    \begin{subfigure}[c]{1\columnwidth}
        \centering
        \includegraphics[width=\linewidth]{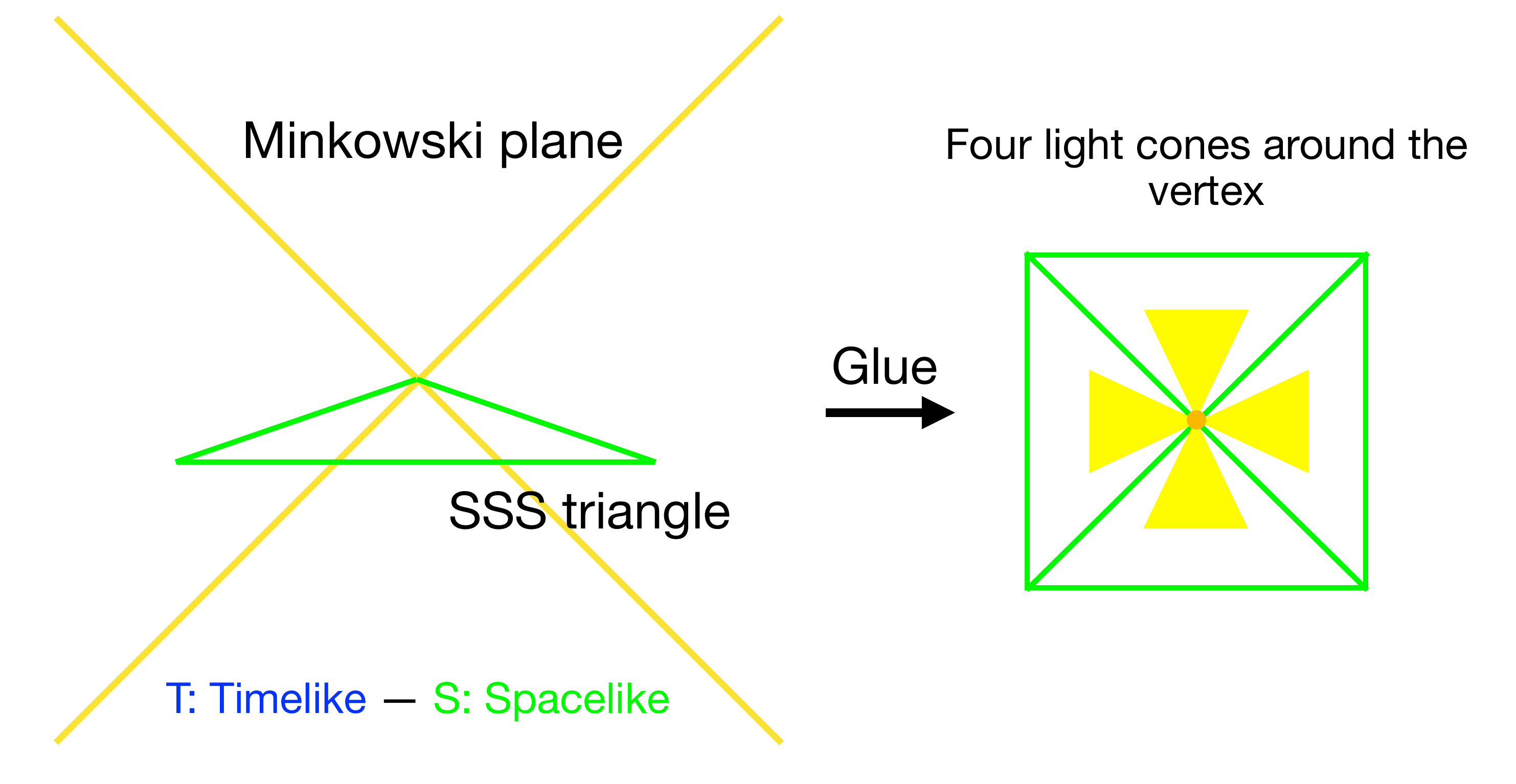}
        \caption{Triangles in the Minkowski plane with all edges being spacelike always have two vertices without light cones and one with a light cone. Gluing four such triangles around the vertices with light cones results in a triangulation with an irregular light cone structure around the bulk vertex: there are \emph{four} light cones there.}
        \label{fig:crotch}
    \end{subfigure}
    \hfill
    \begin{subfigure}[c]{0.5\columnwidth}
        \centering
        \includegraphics[width=\linewidth]{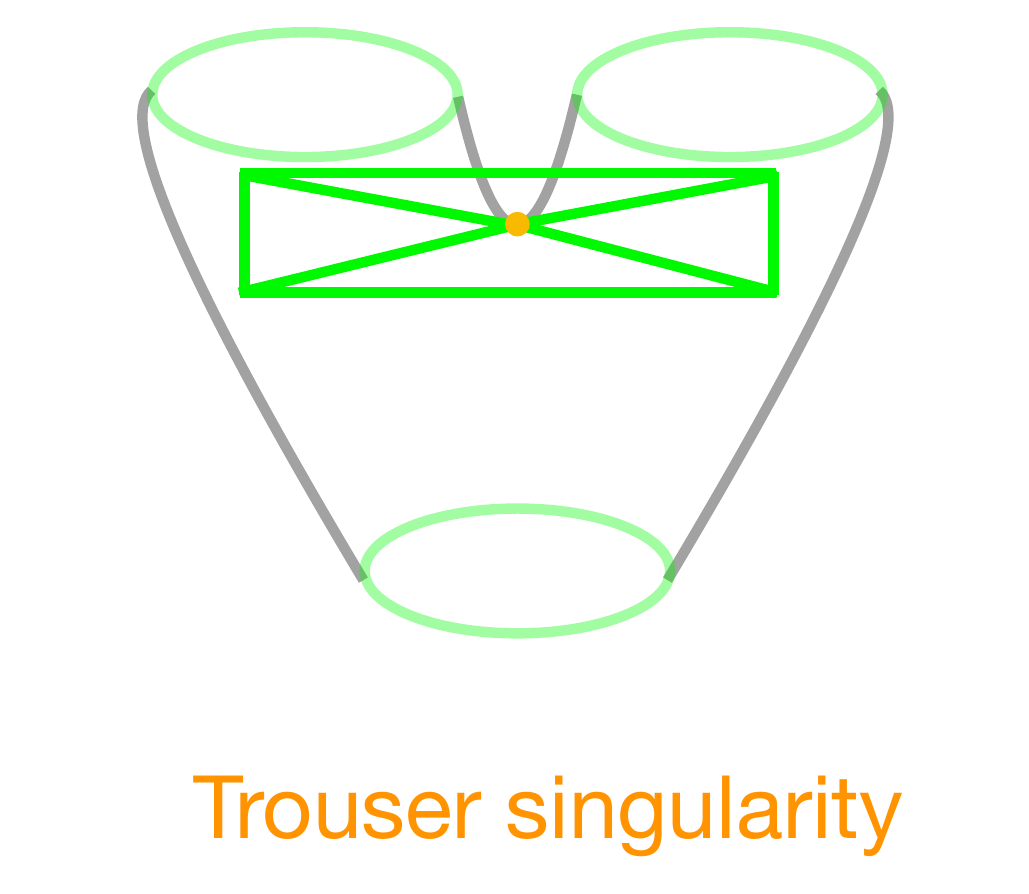}
        \caption{Example of the crotch causal singularity on a trouser spacetime, and how it can be triangulated.}
        \label{fig:trouser}   
    \end{subfigure}
    \caption{A \emph{trouser} Lorentzian spacetime has a light-conical singularity on the crotch, where there is an irregular light cone structure: there are \emph{four} light cones there, instead of the regular two. The crotch can be understood as a point in which the topology of space slices changes: it goes from being one circle to two disconnected circles. More generally, trouser-like light-conical singularities have \emph{more} than two light cones attached to them.}
    \label{fig:trouserlike}
\end{figure}

\begin{figure}[h!]
    \centering
    \begin{subfigure}[c]{1\columnwidth}
        \centering
        \includegraphics[width=\linewidth]{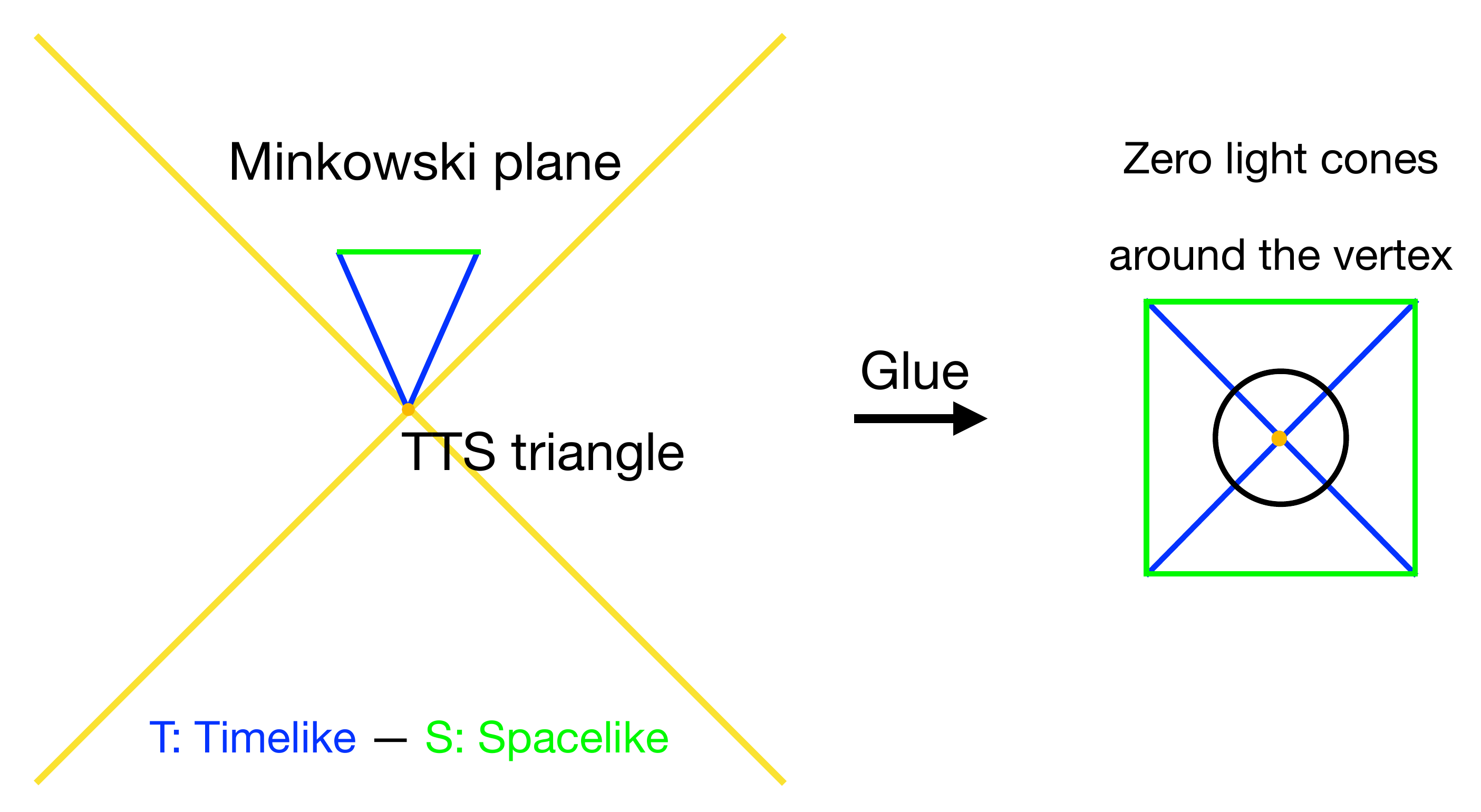}
        \caption{Triangles in the Minkowski plane with two edges being timelike and one spacelike always have zero light rays at the vertex shared by the timelike edges, and one light \emph{ray} attached to the vertices that are shared by one timelike and one spacelike edge. Gluing four such triangles around the vertices with no light rays results in a triangulation with an irregular light cone structure around the bulk vertex: there are \emph{zero} light cones there.}
        \label{fig:yarmulke_tip}
    \end{subfigure}
    \hfill
    \begin{subfigure}[c]{0.5\columnwidth}
        \centering
        \includegraphics[width=\linewidth]{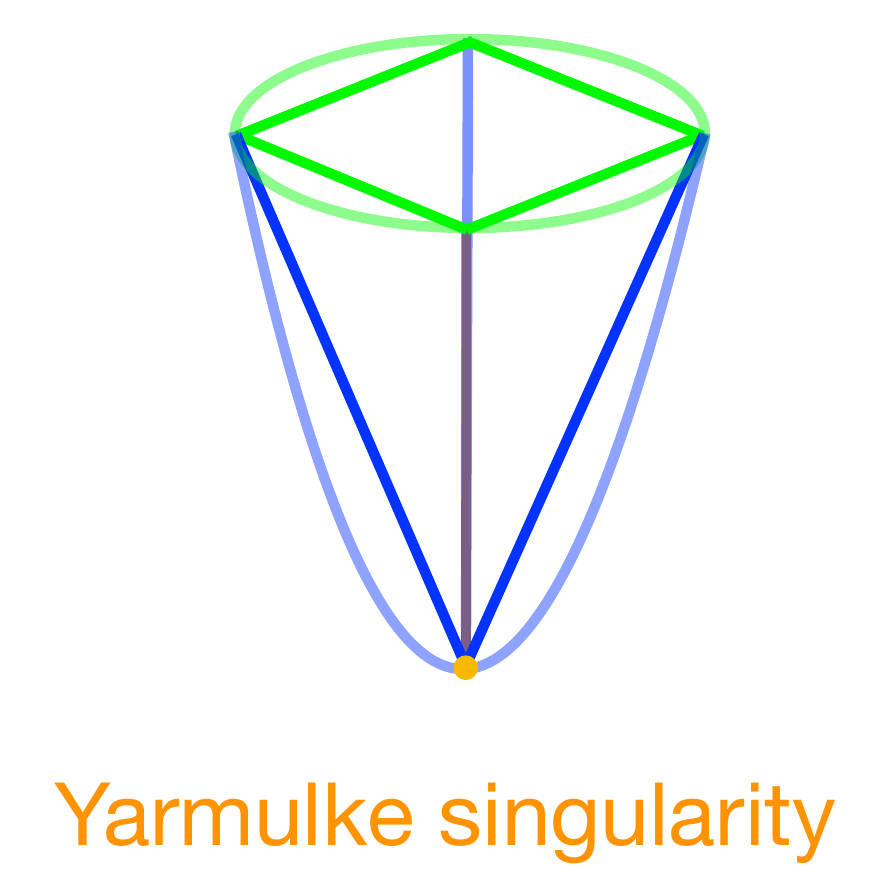}
        \caption{Example of the causal singularity on a yarmulke spacetime, and how it can be triangulated.}
        \label{fig:yarmulke}   
    \end{subfigure}
    \caption{A \emph{yarmulke} Lorentzian spacetime has a light-conical singularity on the tip, where there is an irregular light cone structure: there are \emph{zero} light cones there, instead of the regular two. The tip can be understood as a point in which the topology of space slices changes: it goes from being a point to a circle. More generally, yarmulke-like light-conical singularities have \emph{less} than two light cones attached to them.}
    \label{fig:yarmulkelike}
\end{figure}

As an example, consider a two dimensional triangulation made of four triangles meeting at a vertex —\emph{cf.} FIGS.~\ref{fig:trouserlike} and \ref{fig:yarmulkelike}.  We consider the subspace of geometries where all bulk edges have length squared $a^s$ and all boundary edges have length squared $b^s$. Within this subspace let us consider two cases:
\begin{enumerate}
    \item $a^s>0$ and $b^s>0$, \emph{i.e.} all edges are spacelike. Then, the triangle inequalities require $b^s>4a^s$. An embedding of such a triangle is shown in FIG.~\ref{fig:crotch}. As can be seen in the figure, the angle opposite the longest edge, that is, the $b$-edge, contains a light cone. In our triangulation we have four triangles meeting at the bulk vertex, which is opposite a $b$-edge in each triangle. We therefore have four light cones instead of two at the bulk vertex.

    \item $a^s<0$ and $b^s>0$, \emph{i.e.} the $a$-edges are timelike and the $b$-edges are spacelike. All the directions emanating from the vertex opposite a $b$-edge are timelike. Gluing four of such triangles around this vertex, we have still only timelike directions. We thus have a bulk vertex with zero light cones attached to it. The situation is shown in FIG.~\ref{fig:yarmulke_tip}
\end{enumerate}

Naturally, the scenarios above imply that the triangulation cannot be embedded in flat Minkowski spacetime. But such singularities go beyond simply having a non-zero deficit angle (corresponding to a conical singularity): The causal structure around the bulk point is irregular: it has either too many light cones or too little. 

We will refer to this type of singularities as \emph{light-conical singularities}. Above we considered the two-dimensional case, but similarly to the deficit angle light-conical singularities can be also generalized to higher dimensions: given a bone we consider the projection of the simplices around the bone to the plane orthogonal to the bone. If the bone is spacelike, the plane onto which we project is Lorentzian, and we can count the light ray crossings included in the triangles resulting from the projection. We have a light-conical singularity along the bone, if the number of light ray crossings differs from four.\footnote{We remark that light-conical singularities also have continuum counter-parts with important consequences to quantum gravity, independent of the simplicial approach \cite{Marolf:2022ybi,Colin-Ellerin:2020mva,Marolf:2020rpm,Colin-Ellerin:2021jev,Jacobson:2022jir,Banihashemi:2022jys}.}

In two-dimensions light-conical singularities have a very close relation with topology change. Indeed, as illustrated in FIG. \ref{fig:trouserlike} and FIG. \ref{fig:yarmulkelike} the vertex with an \emph{irregular light cone} structure works as a discrete analogue of Morse points in a topology changing spacetime \cite{Louko:1995jw, Dowker:1999wu,Marolf:2020rpm}, referred to as trouser and yarmulke spacetime respectively. Also in higher dimensions we will refer to bones  with a number of light ray crossings less than four as yarmulke type and to bones with a number greater than four as trouser type.

Remarkably the Regge action responds to light-conical singularities with an imaginary contribution: Each bone with an area $A_b$ contributes an imaginary part\footnote{Consult \cite{Neiman:2013ap,Neiman:2013lxa,Bodendorfer:2013hla,Colin-Ellerin:2020mva} for continuum versions of this result.} \cite{Sorkin:2019llw}
\begin{equation}
    -\imath\left(1-\frac{\mathcal N_c}{4}\right)\frac{A_b}{4G},
    \label{eq:ImS}
\end{equation}
where $\mathcal N_c$ is the number of light ray crossings at the bone. Note that in our units (with $8\pi G=1$), we have $\frac{A_b}{4G}=2\pi A_b$.

Thus light-conical singularities are either enhanced or suppressed in the path integral. With the global sign choice made in (\ref{eq:ImS}) trouser-like singularities are exponentially suppressed and yarmulke-like ones are enhanced. We note however, that there are ambiguities for the sign due to branch-cuts, which come with the light-conical singularities \cite{Asante:2021phx}.  Thus it is the choice of integration contour along these branch-cuts which determines these signs.\footnote{This ambiguity also appears in a vast range \cite{HalliwellContours,Louko:1995jw} of works in continuum general relativity, \emph{e.g.} on  minisupespace cosmology \cite{HalliwellContours} or real-time topology change \cite{Louko:1995jw}. Several works \cite{HalliwellContours,Louko:1995jw,Witten:2021nzp,Kontsevich:2021dmb} have argued that the choice made in \eqref{eq:ImS} is the right one, if one is to make sense of the matter (or graviton) path integral over these kind of spacetimes. It is also the choice that gives sensible answers (positive entropies) in the thermodynamic calculations of \cite{Marolf:2022ybi}, \cite{Colin-Ellerin:2020mva,Marolf:2020rpm,Colin-Ellerin:2021jev} and \cite{Jacobson:2022jir,Banihashemi:2022jys} —see also \cite{Dittrich:2024awu} for a Regge model of the latter work(s). \label{fnote:Halliwell-Hartle_ambiguity}}

At first sight, light-conical singularities might seem like an unwanted feature, especially if they can be exponentially enhanced in the path integral.

However, elaborating on what was mentioned in the introduction, recent work has shown that both trouser- and yarmulke-like singularities might play a fundamental role in the Lorentzian path integral, especially in the context of thermodynamics (without Euclidianization) \cite{Marolf:2022ybi} and entropy calculations \cite{Marolf:2022ybi,Colin-Ellerin:2020mva,Jacobson:2022jir,Banihashemi:2022jys,Dittrich:2024awu, Padua-Arguelles:2025koj}. However, in most of these calculations the light-conical singularity appears in a surface that is topologically special, whereas as suggested by the example above, in the Regge path integral they can be present or not in generic bones, depending on the configuration. Thus, it is important to understand their impact and potential role in quantum Regge calculus.

There have been few works which explored this question: the works \cite{Dittrich:2021gww, Asante:2021phx} considered a cosmological calculation, in the context of the no-boundary wave function, where light-conical singularities are not expected \emph{a priori} from a continuum point of view and compared the results when including/excluding light cone irregular configurations. In this case, the inclusion of the light cone irregular configurations (with an exponentially suppressed amplitude) results in the most physically reasonable scenario and was in better accord with expectations from the minisuperspace continuum theory \cite{Feldbrugge:2017kzv}. Indeed, one can argue \cite{Dittrich:2024awu} that the light cone irregular configurations with the accompanying branch-cuts for the Regge action are a stand-in for the essential singularity occurring for the continuum action\footnote{The action obtained after integrating out the spatial metric degrees of freedom has an essential singularity at vanishing lapse.} for vanishing lapse. The ambiguity of how to navigate the branch cut in the discrete case mirrors the ambiguity of how to navigate the essential singularity in the continuum case.

Somewhat in contrast,
the work \cite{Asante:2021zzh}, did a calculation in a framework closely related to QRC, effective spinfoams, and concluded that their inclusion (with an exponentially enhanced amplitude) led to high deviations from semiclassical expectation values. 

On the other hand \cite{Dittrich:2024awu,Padua-Arguelles:2025koj} discuss Regge triangulations for de-Sitter and back hole spacetimes, respectively, and find that light-conical singularities (appearing on the horizon) need to be included with an exponentially enhanced amplitude in order to produce the expected entropies.

Now, these calculations where performed in discrete version of minisuperspace, \emph{i.e.} they were carried out in a context in which several edge lengths of the triangulations were set to be equal, thereby reducing the number of variables. This is indeed what we did in the example above, by reducing, in a particular way, to the two variables $a^s$ and $b^s$.

So a natural question is whether Regge histories with an irregular light cone structure are actually generic when one considers all possible edge-length assignments, specially when taking the continuum limit. It could be, after all, that in such context they have zero (Lebesgue) measure and that what one was seeing in these works was an artifact of minisuperspace.

Let us use the algorithm above to address this issue.

\subsection{Abundance of light-conical singularities \label{ssec:simulations}}

For this purpose we will consider four-dimensional triangulations, relevant to different refinement schemes for the continuum limit, and apply our the algorithm to generate a large number of histories. For each bulk bone in the triangulation, we will then count how many configurations exhibit $\mathcal{N}_c$ light ray crossings, for all possible values of $\mathcal{N}_c$; \emph{i.e.}, we will construct a light ray crossing histogram for every bone.

\subsubsection{$\Delta_n$ \label{sssec:Delta_n}}

The first class of triangulations we will consider consist of $n$ four-simplices glued around a single triangle and shall be referred to as $\Delta_n$, for each $n$.

We will assume that the bulk triangle is equilateral, with edge length $\lambda$ and we will generate geometries for the boundary of $\Delta_n$, \emph{i.e.} the only fixed geometry in the algorithm will be the one of the triangle. To proceed with the algorithm, as described in \S\ref{sec:algorithm}, we embed the triangle in Minkowski spacetime, and randomly generate two extra vertices to define a four-simplex. This four-simplex can be used as the initial fixed geometry. And this process is repeated on each run of the simulation. For the random generation of these two vertices we will use a uniform distribution in the box $[-L,L]^4$ (with the initial triangle centered at the origin).  The algorithm above will create geometrizations for another $(n-2)$ four-simplices by appending vertices to tetrahedra. The final four-simplex  will be constructed via a dihedral construction. For the random sampling required by the dihedral construction we will use a uniform distribution in the space of squared length.  If the triangle inequalities leave the squared edge lengths unbounded, we will bound it by $L^2$.  This means that we have two scales $\lambda$ and $L$ in the simulation and we must analyze how findings depend on their ratio.  

The motivation for this setup is to study the light cone structure of a typical triangle in the `deep' bulk of a triangulation. We say `deep' bulk because, for triangles close to boundaries, it is reasonable to expect that the geometry of their circuit is more tightly constrained, via triangle inequalities, by the fixed geometry in the boundary. Studying several values of $n$ allows us to see how the light cone structure depends on the circuit's length.  Configurations, as described by the large $n$ limit of $\Delta_n$ appear in the barycentric refinement scheme.  This is one method to take the continuum limit \cite{Rocek:1981yd,Beirl:1994st}, in which the number of simplices around every bone in the original triangulation is taken to infinity.

\begin{figure*}
    \centering
    \includegraphics[width=0.75\textwidth]{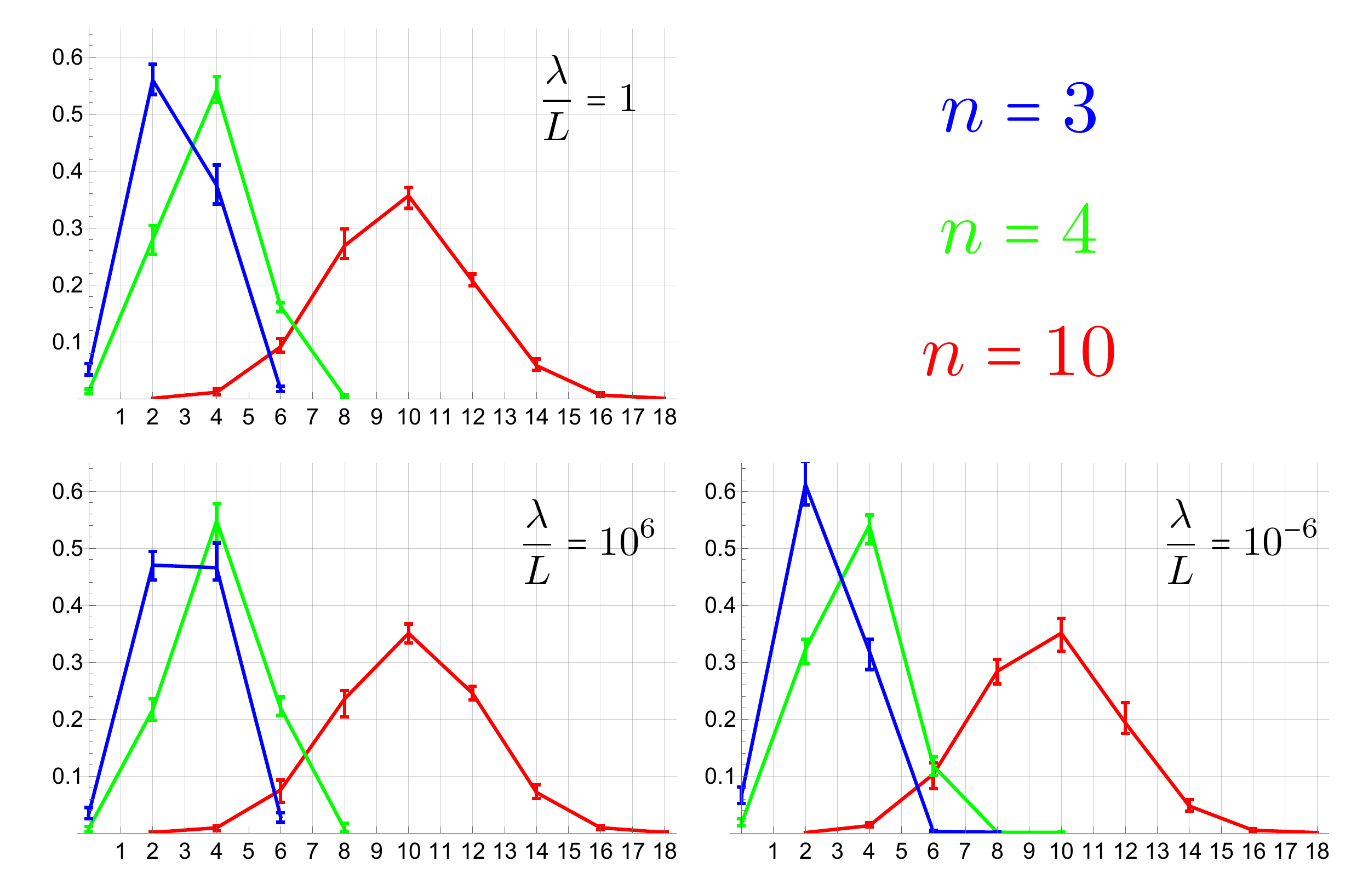}
    \caption{Histograms of the light rays around the central triangle (taken to be equilateral —and thus spacelike) of the triangulation $\Delta_n$, for different values of $n$ and different ratios between the length $\lambda$ of triangle's edges and the scale $L$ used for random samplings involved in the algorithm. Error bars are computed on the basis of having run 10 simulations, each generating 1000 geometrizations.}
    \label{fig:mixed_Delta_n}
\end{figure*}

The results are summarized in FIG. \ref{fig:mixed_Delta_n} (together with some methodology details), where we see that the most typical configuration for $\Delta_n$, regardless of the scales' ratio, is having $n$ light rays emanating from all points inside the central bone. In fact, these histograms are peaked around $n$. For large enough $n$, configurations with the regular number of light rays (or even fewer than that) are extremely rare. This means that for barycentric continuum limit schemes, there is an `entropic' preference for light cone irregular configurations of trouser type.

This raises the question whether a sensible continuum limit can be recovered from Lorentzian Regge gravity, without explicitly forbidding light-conical singular configurations. This should certainly be explored in future work, \emph{e.g.}  via the computation of length expectation values in the path integral. There are at least two mechanisms that can suppress light-conical singular configurations: Firstly, due to the imaginary parts in the Regge action, configurations with light-conical singularities do not satisfy the Regge equations of motion \cite{Sorkin:2019llw,Louko:1995jw,Marolf:2020rpm}. Thus, in the path integral oscillations of light-conical singular configurations will typically lead to destructive interference.  The question then is whether this can help overcome the entropic preference —as an analogy we might remember that for the Euclidean path integral of a non-relativistic particle, the bulk of geometries are continuous everywhere but nowhere differentiable, and nonetheless the classical limit is sensible. 


Secondly, we note that for sufficiently large $n$ light-conical singularities of trouser type entropically dominate over those of yarmulke type. This suggests to resolve the sign ambiguity mentioned in footnote \ref{fnote:Halliwell-Hartle_ambiguity} (\emph{i.e.} to choose the gravitational integration contour along branch cuts accordingly) as done in \eqref{eq:ImS}, which is also following the choice of sign in \cite{Louko:1995jw} by Louko and Sorkin. With this choice one suppresses exponentially trousers and exponentially enhances yarmulkes.\footnote{An example where one has exponentially enhanced yarmulke configurations with unbounded integration domain and one nevertheless obtains sensible results can be found in \cite{Dittrich:2024awu}.} This choice helps also to ensure convergent matter fluctuations \cite{Louko:1995jw,Kontsevich:2021dmb}.

If such a mechanism suppressing trousers is absent, one might have to choose a refinement procedure in which the number of simplices around a bone remains bounded —see \emph{e.g.} the next section. Another possibility is to modify the action, so that there is an even stronger suppression of light-conical singularities.\footnote{This can be justified by the program of perfect actions, which aims to construct discrete actions which mirror perfectly continuum physics \cite{Hasenfratz:1993sp, Bahr:2009qc}. }

Above we found that the distribution in the number of light rays emanating from the central bone in $\Delta_n$ is peaked around $n$. Indeed, projected wedges can either include zero, one, or two light rays. We thus seem to have a `balanced' distribution between these choices, in the sense that neither wedges with zero nor wedges with two light rays dominate.  

Let us now consider a modified set of configurations. Here we only allow 4-simplices that contain only spacelike tetrahedra (and therefore triangles and edges). We note that the so-called EPRL spin foams \cite{Engle:2007wy} were in their original formulation\footnote{The models have been extended to allow also for timelike tetrahedra, triangles and edges \cite{Conrady:2010kc}. But for technical reasons numerical tools developed so far for these models  often restrict to the case of having only spacelike tetrahedra \cite{Dona:2022yyn}.} restricted in this way. Given the fact that Regge geometries appear in the semiclassical limit of spin foams, see \emph{e.g.} \cite{Barrett:2009gg}, this simulation provides some insight into the light cone structure of spin foams in that regime.x

\begin{figure*}
    \centering
    \includegraphics[width=0.75\textwidth]{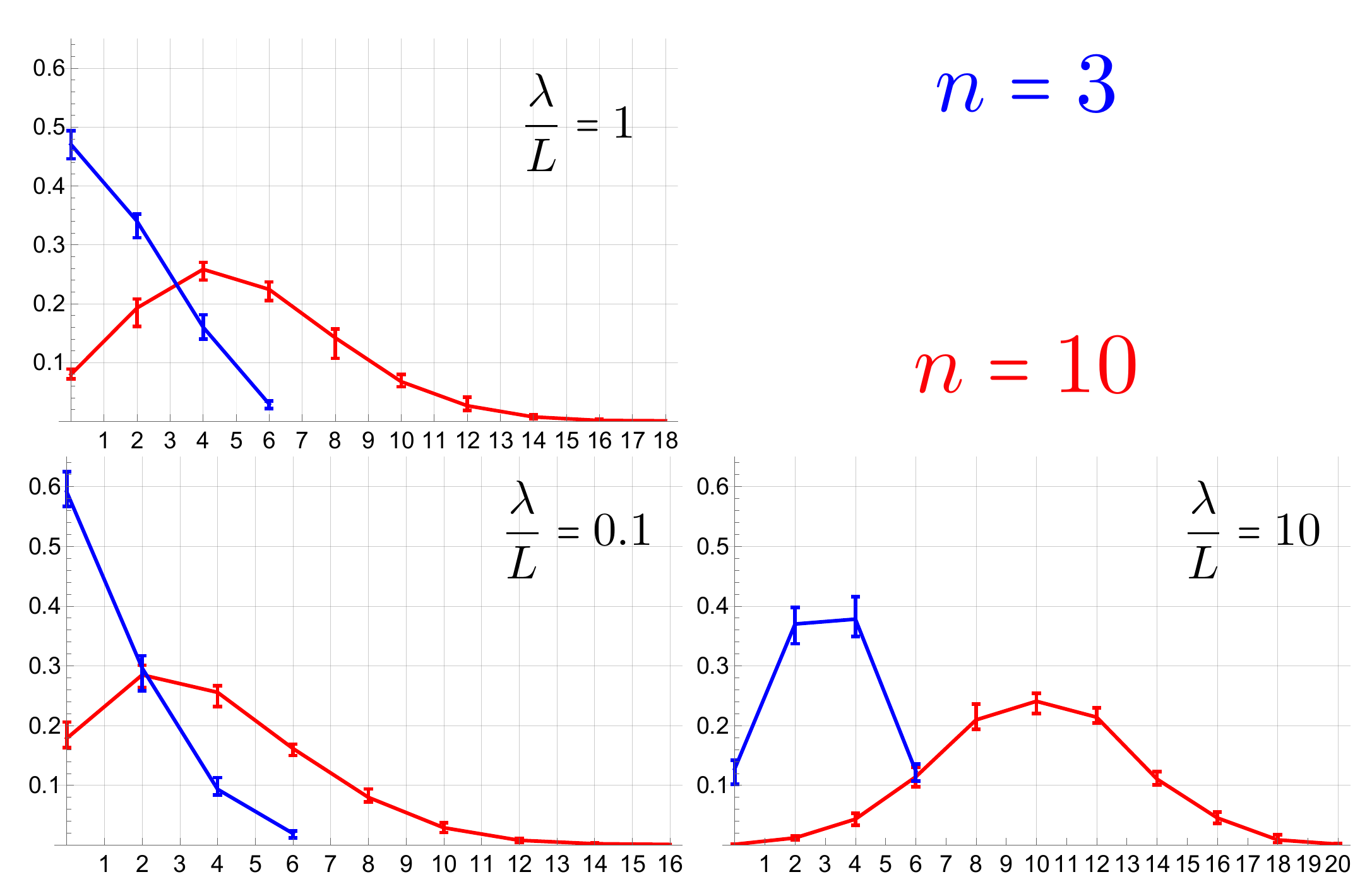}
    \caption{Histograms for the number of light rays around the central triangle (taken to be equilateral —and thus spacelike) in the triangulation $\Delta_n$, for different values of $n$ and different ratios between the length $\lambda$ of triangle's edges and the scale $L$ used for random samplings involved in the algorithm. Error bars are computed on the basis of having run 10 simulations, each generating 1000 geometrizations. Only geometrizations in which all proper subsimplices were spacelike were considered.
    }
    \label{fig:spacelike_Delta_n}
\end{figure*}

The simulation results are depicted in FIG. \ref{fig:spacelike_Delta_n}. We see that the peak of the histograms depends (much more heavily than in the previous case) on the ratio $\lambda/L$. Additionally, for larger $L$ the peak of the distribution tends to go to light-conical singularities of yarmulke type.  We thus see that restricting the configuration space by allowing only spacelike tetrahedra can heavily influence the abundance of light-conical singularities.

\subsubsection{Subdivided hypercube}

\begin{figure}
    \centering
    \includegraphics[width=0.25\textwidth]{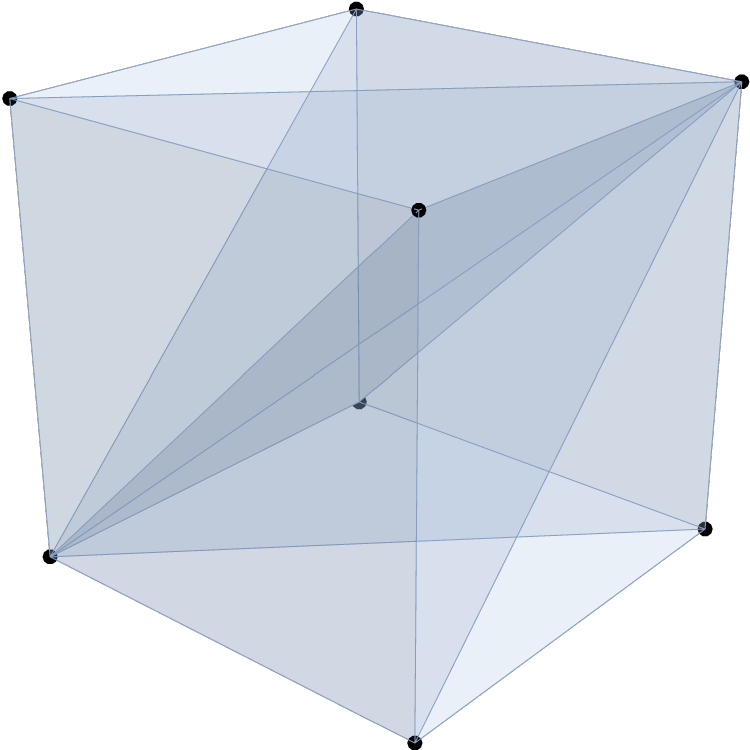}
    \caption{Standard triangulation of a 3-cube. A particular gluing of eight of these triangulations makes up the boundary of the triangulation used for the hypercube simulation. The full triangulation connects all boundary vertices  with one vertex in the bulk.}
    \label{fig:standard_hypercube_triangulation_3D}
\end{figure}

Given the possible drawbacks of the barycentric subdivision scheme to approach the continuum limit, we now explore a triangulation relevant to another refinement scheme that has been considered in the literature \cite{Rocek:1982tj,Dittrich:2021kzs,Dittrich:2022yoo}. Here we consider the triangulation of a hypercube constructed as follows: We first triangulate the boundary of the hypercube, which consists of eight 3-cubes, by using the so-called ``standard'' triangulation  shown in FIG.~\ref{fig:standard_hypercube_triangulation_3D}. Then we connect each of these boundary tetrahedra to a vertex in the bulk of the hypercube. This defines the connectivity of a triangulation of a hypercube, which has one bulk vertex. 

The triangulated hypercubes can then be glued to each other, obtaining a triangulation of a hypercube (or hyper torus) with an arbitrarily large number of simplices. We will apply  the realizability algorithm to one hypercube to (statistically) study the light cone structure of histories for this triangulation.

We will consider all edge-length variables of this triangulation to be dynamical, including those of the boundary edges. This is meant to explore the typical light cone structure in one basic hypercube deep in the bulk of the large triangulation we just described.

\begin{figure}
    \centering
    \includegraphics[width=0.5\textwidth]{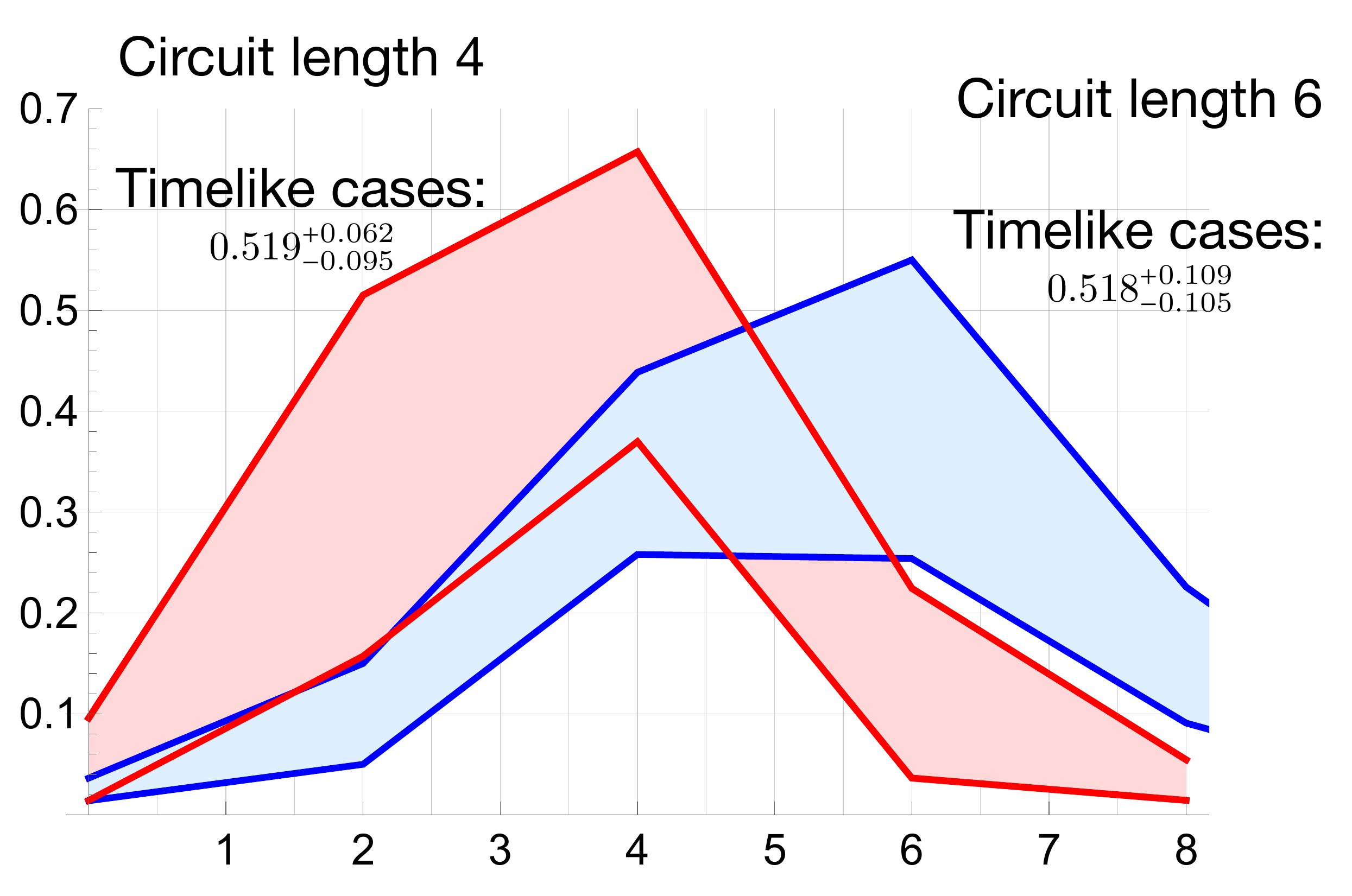}
    \caption{Bands generated by the upper and lower bounds of light ray histograms for all spacelike bones of circuit length four and six in the hypercube triangulation. Also displayed are the average portion of configurations where the bones are timelike, along with the upper and lower bounds found in the simulations for these portions. The number of bulk bones in the hyper-cube with circuit length four (six) is forty-eight (sixteen). The simulation consisted of 500 geometrizations.}
    \label{fig:hypercube_simulation}
\end{figure}

The simulation results are shown in FIG.~\ref{fig:hypercube_simulation}. We see that the conclusions of the previous subsection still hold: The most typical configurations have bulk bones with a number of light \emph{rays} equal to the length of the bone's circuit. The triangulation includes  sixteen bulk bones of circuit length six. Thus, light cone irregular configurations have a very high probability.

\section{Light conical singularities for large bulk edges}
\label{Sec:LargeBulk}

In the previous section we considered the frequency of different types of light-conical singularities in configuration space. In order to be able to generate a sampling of the configuration space, we had to restrict to triangulations with finite edge lengths. \emph{E.g.} when sampling for new vertex coordinates we restricted the new vertex to be inside a box of a certain size.

Here we will consider configurations where we can make the bulk edges arbitrarily large while keeping boundary edges fixed, and investigate their behavior in the limit of infinitely large bulk edges. It was found in \cite{Borissova:2024pfq,Borissova:2024txs} that such configurations can lead to light-conical singularities with unbounded support in the bulk length configuration space.  Here we generalize and systemize the findings of \cite{Borissova:2024pfq,Borissova:2024txs}, showing that both yarmulke- and trouser-like light-conical singularities can appear with unbounded support.

The issue with the appearance of such configurations is that in the case of exponential enhancement of such light-conical singularities, the unbounded support might lead to a divergent path integral. As we have already noted, whether these singularities are enhanced or suppressed depends on the choice of integration contour. Presumably one could choose it such that one suppresses configurations that have both yarmulkes and trousers.\footnote{Assuming that there are no topological obstructions to do so. We will encounter below examples where a given configuration features both yarmulkes and trousers. The total imaginary part of the action then determines on which side of the branch cut such configurations are enhanced or suppressed.} We wish to also point out, that even if a light-conical singularity has unbounded support and is exponentially enhanced, the resulting path integral might be finite. An example for this can be found in \cite{Dittrich:2024awu}, where a yarmulke singularity appears with unbounded support (\emph{i.e.} for arbitrarily large size of the singularity), and were the path integral with exponential enhanced amplitude does converge. This result however crucially depends on the order of integration: One has to first integrate over the variable(s) which do not include the size of the singularity. In the case of \cite{Dittrich:2024awu} this leads to a destructive interference, which results in a vanishing amplitude for sizes of the horizon larger than a certain critical value. A crucial question for future research will be whether a similar mechanism is at work for more general situations —consult \cite{Marolf:2022ybi} for a closely related phenomenon. 

Three- and four-dimensional triangulations show different behavior; we therefore will discuss these cases separately. 
~\\~\\
\emph{Three-dimensional triangulations:}\\
Starting with three-dimensional triangulations, we first consider configurations which were named ``spine configurations" in \cite{Borissova:2024pfq,Borissova:2024txs}. These consist of $N$ tetrahedra that share one edge. There are no bulk vertices and no further bulk edges. For Lorentzian signature triangulations there are boundary data for which the length of the bulk edge can become arbitrarily large.\footnote{For Euclidean triangulations the triangle inequality prevents such cases: for each bulk triangle the bulk edge length has to be smaller than the sum of the lengths of the other two edges, both of which are in the boundary. In contrast, the Lorentzian triangle inequalities may demand, \emph{e.g.} from a timelike triangle with only spacelike edges, that there is one edge whose length is larger than the sum of the length of the other two edges.}

For a three-dimensional triangulation, bones are edges, and we can have only light-conical singularities around spacelike bones, that is, spacelike edges. Consider a Lorentzian tetrahedron with one edge becoming very large, whereas the other edge lengths are fixed. The dihedral angle at the very long edge has to be a ``thin Lorentzian angle'' —\emph{i.e.} its projected triangle does not contain a light ray crossing. In fact, the angle tends to zero in the limit where the large edge length is going to infinity —see \cite{Borissova:2024pfq} for a detailed derivation.  Thus, if we have $N$ tetrahedra glued around a very long spacelike edge, the full angle around this edge contains only thin Lorentzian angles, and therefore no light ray crossing. We therefore have yarmulke like light-conical singularities. A more detailed analysis \cite{Borissova:2024pfq} shows that the leading order of the action is determined by this yarmulke light-conical singularity and thus given by
\begin{align}
S=  2\pi \imath l + {\cal O}\left(\log l\right),
\end{align}
where $l$ is the length of the bulk edge.

Next we consider so-called ``spike configurations" in three-dimensional triangulations. We construct these by starting with the triangulation of a 2-sphere, from which we construct a (Lorentzian) triangulation of a 3-ball. To do so, we place one vertex inside the ball and connect all vertices in the boundary to this bulk vertex. We likewise associate a bulk triangle to each edge in the boundary and a tetrahedron to each triangle in the boundary, with all bulk edges, bulk triangles and tetrahedra sharing the bulk vertex. We consider the case where we have only spacelike edges and additively scale all bulk edges to be large, that is, $l_b\rightarrow l_b+\lambda$ for all bulk edges $b$, with $\lambda$ becoming large. At leading order this is equivalent to setting $l_b=\lambda$. 

Consider a Lorentzian tetrahedron from this triangulation, that is, a Lorentzian tetrahedron with one `base' triangle with bounded edge lengths and three spacelike edges whose edge lengths become homogeneously large. To satisfy the Lorentzian triangle inequalities, the `base' triangle needs to be timelike \cite{Borissova:2024pfq}. The boundary triangulation therefore consists of timelike triangles.

In the limit of infinite edge length the three infinitely long edges meet at infinity and can therefore be considered as parallel. Indeed, in the limit, the dihedral angle at such an edge becomes equal to the two-dimensional angle at the vertex in the `base' triangle to which the infinitely long edge is attached.  One can speak of a holographic projection \cite{Riello:2013bzw, Borissova:2024pfq,Borissova:2024pfq}, as to leading order in the large edge lengths the 3D action is determined by the 2D action of the boundary triangulation. The boundary triangulation has the topology of a sphere, and due to the (Lorentzian) Gauss-Bonnet theorem, its action is given by $ 4\pi\imath$. For the 3D action we have to weight with the lengths of the bulk edges, and we obtain (see \cite{Borissova:2024pfq} for a detailed derivation)
\begin{align}
S=  4\pi \imath l + {\cal O}\left(l^0\right) \quad ,
\end{align}
where $l$ is the length of the bulk edges.

Because of what we said above, a given bulk edge  is light-conical singular (to leading order), if the associated boundary vertex is light-conical singular with respect to the Lorentzian 2D triangulation. Yarmulke/trouser singularities in the boundary lead to yarmulke/trouser singularities in the bulk. Overall (with all bulk edges having equal length), because of the Lorentzian Gauss-Bonnet theorem for a sphere, the contributions from the yarmulke singularities over-compensate the ones from the trouser singularities.

To provide examples of Lorentzian 2D triangulations with yarmulke like and trouser-like singularities we consider (slight generalizations of) a triangulation constructed in \cite{Dittrich:2024awu}, which discussed a minisuperspace model of 3D de Sitter space:
\begin{itemize}
\item Start with a `triangulation' of $S^1$, consisting of $N\geq 2$ edges of spacelike squared length $s_a>0$.
\item We take this triangulation of $S^1$ as a boundary of a Lorentzian disk, which we triangulate by connecting a bulk vertex with the $N$ boundary vertices via edges of squared length $s_b$. These edges can be null, timelike, or spacelike. The Lorentzian triangle inequality implies $s_b<s_a/4$.
\item Taking two such triangulations of the disk we glue them along the $S^1$-boundary and obtain a Lorentzian triangulation of the 2-sphere.  All triangles in this triangulation have the same geometry and are timelike, having one edge of squared length $s_a>0$, and two edges of squared length $s_b<s_a/4$.
\end{itemize}

We consider the case where $s_b$ is spacelike and the case where $s_b$ is timelike. We will comment on the null case in  Section \ref{sec:chrono}. 
\begin{itemize}
\item For the case where $s_b$ is spacelike, $0<s_b<s_a/4$, consider the vertices where the edges of $s_a$ and $s_b$ meet. The angle in a given triangle is a thin Lorentzian angle between two spacelike edges, which does not contain a light ray crossing. The full angle around such a vertex therefore also does not contain any light ray crossing and thus features a yarmulke like singularity. Next, consider the vertices which are shared by only $s_b$-edges. The angle in a given triangle is a thick Lorentzian angle containing two light ray crossing and thus one full light cone. The full angle around such a vertex contains therefore $N$ light cones. These vertices are therefore light cone regular for $N=2$, but feature a trouser like singularity for $N>2$. The action evaluates to $\imath (2\pi N- 2( N-2)\pi)= \imath 4\pi$.
\item If $s_b$ is timelike, the angle between an $s_a$- and  an $s_b$-edge now includes one light ray crossing. Four triangles meet around a vertex shared by $s_a$ and $s_b$ edges, such a vertex is therefore light cone regular. At the vertices shared by only $s_b$ edges the angles do not include a light ray crossing. We therefore have two vertices with  yarmulke singularities. The action again evaluates to $\imath 4\pi$.
\end{itemize}

Another  triangulation of the Lorentzian 2D sphere that includes  yarmulke and trouser singularities can be constructed as the boundary of a 3D triangulation as follows: Start with a  tetrahedron that has only spacelike triangles. This tetrahedron can have either Lorentzian or Euclidean signature, as in the end it is its boundary that we will care about. In the latter case, we can take it as equilateral; in the former case we can embed the entire construction into a 3D Minkowski space.  Glue a tetrahedron with a spacelike base triangle and three timelike edges to each of the triangles of the initial tetrahedron. We then consider the boundary of this triangulation, which consists of only timelike triangles, as the triangulation of the sphere. It has four vertices with six instead of four light ray crossings (trousers) and four vertices with no light ray crossings (yarmulkes). The contribution of the yarmulke singularities overcompensates those of the trouser ones.

See \cite{Borissova:2024pfq} for a discussion of other cases of spike configurations, including ones in which there are spacelike and timelike large bulk edges. 
For all the cases discussed in \cite{Borissova:2024pfq}, except the case of spike configurations with all bulk edges spacelike, large spacelike edges lead to yarmulke singularities only. For the case of a spike configuration with all bulk edges spacelike, we have shown that trouser singularities can appear. But their  contribution to the action is overcompensated by the contribution resulting from the yarmulke singularities —in the case that all bulk edges have the same length to leading order. 

~\\
\emph{Four-dimensional triangulations:}\\
We will see that the situation for four-dimensional Lorentzian triangulations differs from the one for three-dimensional triangulations. 

Four-dimensional spine configurations are defined as configurations where $N$ four-simplices share a very large bulk edge. In four dimensions, the triangles are bones, so triangles support distributional curvature. In a spine configuration the     `bulk' triangles have one very large edge and two edges with small lengths. This is only possible for timelike triangles. All bulk triangles are therefore timelike and come with Euclidean deficit angles, which by definition cannot be light-conical singular. If we have a large timelike bulk edge, the bulk triangles which share this edge needs also to be timelike.

However, for four-dimensional spike configurations we do encounter light-conical singularities. These can be constructed by starting with the triangulation of a 3-sphere. Taking this 3-sphere triangulation as the boundary of a 4-ball, we obtain a four-dimensional triangulation by placing one vertex inside this ball and connecting all boundary vertices to the bulk vertex. We associate bulk triangles to boundary edges, bulk tetrahedra to boundary triangles, and four-simplices to boundary tetrahedra one-to-one. We consider the case in which all bulk edges are spacelike and scaling $l_b\rightarrow l_b+\lambda$ with $\lambda$ becoming large, for the bulk edge lengths. The Lorentzian generalized triangle inequalities impose that for this limit to be allowed, the boundary tetrahedra need to be timelike \cite{Borissova:2024txs}.  We can therefore assume a three-dimensional Lorentzian boundary triangulation. 

For the same reason as for three-dimensional spike configurations (with all bulk edges having the same signature and homogeneous scaling), a holographic projection is also at work in the four-dimensional case: the deficit angle at a given bulk triangle is, in the limit of infinite large bulk edges, equal to the three-dimensional deficit angle at the boundary edge to which the given bulk triangle is attached. To leading order, the four-dimensional Regge action is determined by the Regge action of the  boundary triangulation \cite{Borissova:2024txs}
\begin{align}
S= \frac{l}{2} S^{\rm 3D} + {\cal O}\left(l^0\right) \quad ,
\end{align}
with $l$ denoting the length of the bulk edges.
 Unlike the two-dimensional Regge action, the three-dimensional Regge action is not determined by a topological invariant. Indeed,  the three-dimensional Regge action for a 3-sphere can be real or have a positive or negative imaginary part.

To see this, we construct a family of triangulations for the 3-sphere, which can then be completed to spike configurations, or more precisely to triangulations of the 4-ball, as described above. This family of 3-sphere triangulations generalizes the 2-sphere triangulations we considered above:
\begin{itemize}
\item Start with the triangulation of a 2-sphere. We use a regular triangulation in which all edges have the same edge length squared $s_a>0$. That is, we consider the triangulation of the sphere with two triangles (where all edges are pairwise identified), or the boundary of a tetrahedron, or the boundary of an octahedron, or the boundary of an icosahedron. We note that a vertex in these triangulations is shared by two, three, four, and five triangles, respectively.
\item We take the triangulation of the 2-sphere as the boundary of a Lorentzian 3-ball, and triangulate this 3-ball by introducing one bulk vertex and connecting this bulk vertex with all vertices in the boundary triangulation. We assume that all additional edges have the same edge length squared $s_b$. To satisfy the Lorentzian generalized triangle inequalities (remember that the tetrahedra need to be timelike to allow for very large edges in the four-dimensional bulk of the spike configuration), we need $s_b< s_a/3$.
\item Taking two such triangulations of the 3-ball we glue them along the 2-sphere boundary. We thus obtain a Lorentzian triangulation of the 3-sphere. Note that all tetrahedra in this triangulation have the same geometry: a given (timelike) tetrahedron has a spacelike `base' triangle with squared edge lengths $(s_a,s_a,s_a)$. This base triangle is connected to the fourth vertex via edges of squared lengths $s_b$, the three remaining triangles have, therefore, squared edge lengths $(s_a,s_b,s_b)$
\end{itemize}

These triangulations of the 3-sphere exhibit three different regimes in which different types of light conical singularities, or no light conical singularities, appear. (We ignore cases of vanishing measure in the edge length configuration space. That is the cases where either the edges of squared length $s_b$ or the triangles with squared lengths $(s_a,s_b,s_b)$ are null.) 
\begin{itemize}
\item The first regime is characterized by the triangles with edge lengths $(s_a,s_b,s_b)$ being spacelike (and thus $s_b>0$). In this case the dihedral angle at the $s_a$-edges is between two spacelike triangles, and is a thin Lorentzian angle, containing no light ray crossing. There are four tetrahedra glued around a given $s_a$-edge, which all include no light ray crossing. These edges are therefore light-conical singular and of yarmulke type. The dihedral angles at the $s_b$-edges are between two spacelike triangles, but this time it is a thick Lorentzian angle, containing two light ray crossings. Depending on the 2-sphere triangulation we started with, a given $s_b$-edge is shared by either two, three, four, or five tetrahedra. Correspondingly, the $s_b$-edges are either light cone regular, or light cone singular of trouser type. 
\item The second regime is characterized by the triangles with edge lengths $(s_a,s_b,s_b)$ being timelike, but the $s_b$-edges being spacelike, $s_b>0$. The dihedral angle at an $s_a$-edge is between a spacelike and a timelike triangle, containing one light ray crossing. As four tetrahedra are glued around an $s_a$-edge, the $s_a$-edges are now light cone regular. But the dihedral angle at an $s_b$-edge is now between two timelike triangles and is a thin Lorentzian angle containing only timelike directions and no light ray crossing. The $s_b$ edges are therefore light-conical singular of yarmulke type.
\item The third regime is characterized by the $s_b$-edges being timelike. The dihedral angles at these edges are therefore Euclidean and by definition light cone regular. The dihedral angles at the $s_a$-edges are, as in the previous case, between a spacelike and a timelike triangle, and contain one light ray crossing. The $s_a$-edges are thus light cone regular also in this regime. (The vertices shared by only $s_b$ edges are light cone irregular, but this does not lead to imaginary terms in the action.) 
\end{itemize}

Thus we have in the first regime yarmulke singularities, and trouser singularities if the $s_b$ edge is shared by more than two tetrahedra. For this regime the squared edge length $s_b$ is bounded by $4 s_a/12 > s_b > 3 s_a/12$. Computing the action one finds that the yarmulke contributions always dominate over the trouser contributions.  The second regime features only light-conical singularities of yarmulke type, whereas in the third regime we do not have light-conical singularities (which by definition are supported on codimension-two simplices), but have vertices which are light cone irregular.

Next, we construct Lorentzian triangulations of the 3-sphere, where we have only light-conical singularities  of trouser type. To this end, consider a generalization of the second type of triangulations of the 2-sphere with trouser-like and yarmulke singularities discussed previously: Start with an equilateral (and therefore Euclidean) four-simplex and glue onto each of the five tetrahedra of this four-simplex a Lorentzian four-simplex with a spacelike equilateral tetrahedron as a base. We take the remaining edges of these Lorentzian four-simplices as timelike and we can choose them to be all of the same length. The boundary of this triangulation does define the triangulation of a 3-sphere with only timelike tetrahedra. 

Within this Lorentzian triangulation of the 3-sphere only spacelike edges can support light-conical singularities. There are 10 spacelike edges, each of which is shared by six tetrahedra. The dihedral angle at a spacelike edge in a given tetrahedra is between a spacelike and a timelike triangle and therefore includes one light ray crossing. Therefore the full angle at a spacelike edge includes 6 light ray crossings, and the deficit angle has an imaginary part  given by $- \imath \pi$. The imaginary part of the 3D Regge action sums up to $- 10\imath\pi l$, where $l$ is the length of the spacelike edges. Note that although the timelike edges are not light-conical singular, the vertices which are shared by only timelike edges are light cone irregular: Surrounding such a vertex by a sphere, all the directions going through the sphere are timelike.

~\\
\emph{Summary:}\\
We considered certain types of 3D and 4D Lorentzian triangulations with large bulk edges. 

3D spine configurations consists of $N$ tetrahedra glued around a very large bulk edge. This large bulk edge carries a yarmulke like singularity. 3D spike configurations have a boundary of spherical topology and one bulk vertex whose adjacent bulk edges are all very large. The 3D Regge action is to leading order (in the length of the bulk edges) given by the 2D action of the boundary. If all bulk edges are spacelike and large, the boundary triangulation has to be Lorentzian. 3D light-conical singularities supported on the bulk edges are `mirrored' by 2D light-conical singularities supported on the boundary vertices of the boundary triangulation. Due to the Gauss-Bonnet theorem, which determines the 2D Regge action, the boundary triangulation has to have yarmulke-like singularities. Trouser-like singularities can appear, but (if all the lengths of the bulk edges are equally large) their contribution is over-compensated by the contributions from the yarmulkes.

4D spine configurations consists of $N$ four-simplices glued around a very large bulk edge. But due to the Lorentzian triangle inequalities all bulk triangles sharing this bulk edge have to be timelike and can therefore not support light-conical singularities. 

4D spike configurations have a boundary which has the topology of a 3-sphere, and a bulk vertex with all adjacent edges having large lengths. If all bulk edges are spacelike and of the same large length, the action of such a 4D configuration is to leading order determined by the action of its Lorentzian boundary triangulation. We can construct Lorentzian triangulations of the 3-sphere with only yarmulke-like light-conical singularities, only trouser-like light-conical, both types of singularities or with no light-conical singularities at all.

\section{Causal irregularities supported on edges and vertices }
\label{Sec:Other}

So far we mainly discussed irregularities in the light cone structure supported on bones, that is simplices of codimension-two.

But causal irregular structures can be also associated to simplices of codimension-three and four —for four-dimensional triangulations. In particular, \cite{Jordan:2013awa} introduced the following two regularity conditions for four-dimenionsinal triangulations:

\begin{itemize}
    \item \textbf{Edge causality} concerns spacelike edges. For a given spacelike edge $e$, one considers, within each 4-simplex $\sigma$ containing $e$, a cut orthogonal to $e$ and centered at its midpoint. This produces a 3D Lorentzian piecewise flat manifold $\Sigma$ with a spherical boundary and a single bulk vertex, given by the midpoint of $e$. The causality condition requires that exactly two light cones emanate from this interior vertex, such that their intersection with the spherical boundary yields two disconnected disks.

    \item \textbf{Vertex causality} pertains to vertices in the four-dimensional triangulation. For a vertex $v$, one considers its star ---the union of all simplices containing $v$. The causality requirement is that the light cones based at $v$, when intersected with the boundary of this star, produce two disconnected three-dimensional balls.
    \end{itemize}

Violations of these causality conditions (if not accompanied by a light-conical singularity, which is supported on bones) do not necessarily imply an imaginary part in the Regge action. Nevertheless it is important to understand in which way they effect the Regge path integral. \cite{Asante:2025qbr} provided a study and classification of vertex causality violations for 3D triangulations.

\section{Chrono-topology and pairwise embeddability}
\label{sec:chrono}

We discussed light cone irregularities supported on codimension-two simplices, as well as co-dimension three and co-dimenion four simplices —the latter in four dimensions. Can there be also irregularities in the causal structure associated to codimension-one simplices? 

For a Euclidean signature triangulation one can always embed two $d$-simplices sharing a $(d-1)$-simplex into $d$-dimensional Euclidean flat space in a way they are actually glued when embedded, without overlapping. If this would be also the case for Lorentzian triangulations, we would not expect causal irregularities associated to codimension-one simplices. We will however see, that this is not the case:\footnote{We remind the reader that the (Lorentzian) generalized triangle inequalities for a given $d$-simplex are equivalent to the embeddability of this simplex into $d$-dimensional (Minkowski) flat space. We also demand that the geometry induced by two neighboring $d$-simplices onto the shared face agrees.} It is possible for two neighboring simplices in a Lorentzian triangulation to be individually realizable, yet admit no non-overlapping embedding in which they remain glued along their shared face. And indeed, we will see that this obstruction is related to a non-standard causal feature.

For this, let us first define precisely what we mean by ``overlapping’’. If two neighboring realizable simplices are embedded in a way they remain glued, then the embeddings meet in a codimension-one subsimplex. The (embedded) edges of said subsimplex span a hyperplane and thus divide the space into two sides. If the two simplices lie on the same side, they overlap; otherwise they do not. In the former case, we say that the simplices have opposite (relative) (spacetime) orientation. 

If every pair of neighboring $d$-simplices in a $d$-dimensional Lorentzian triangulation admits an embedding in which they remain glued and have the same relative orientation, we call the triangulation \emph{pairwise embeddable}.

Note that a pairwise embeddable triangulation does not imply that there is a globally consistent time orientation —\emph{i.e.} a direction for the flow of time. If there is a globally consistent time orientation for a given triangulation (so that each pair of simplices can be embedded into Minkowski space and the required time orientations agree), we name it a pairwise embeddable chrono-topology for this triangulation. 

We should remark that 

\begin{itemize}
\item[(a)] There are pairs of Lorentzian $d$-simplices sharing a $(d-1)$-simplex which are not embeddable with positive relative spacetime orientation —\emph{i.e.} there are triangulations that are not pairwise embeddable. This can only happen if the shared $(d-1)$-simplex is null.\footnote{This in particular illustrates that Euclidean Regge geometries are always pairwise embeddable.}
\item[(b)] There are triangulations which are pairwise embeddable but do not admit a pairwise embeddable chrono-topology.
\item[(c)] Fixing the connectivity of a triangulation but changing the lengths of the edges (in particular from spacelike to timelike or vice versa) can lead to different pairwise embeddable chrono-topologies.
\end{itemize}

\begin{figure}[h!]
    \centering
    \begin{subfigure}[c]{1\columnwidth}
        \centering
        \includegraphics[width=\linewidth]{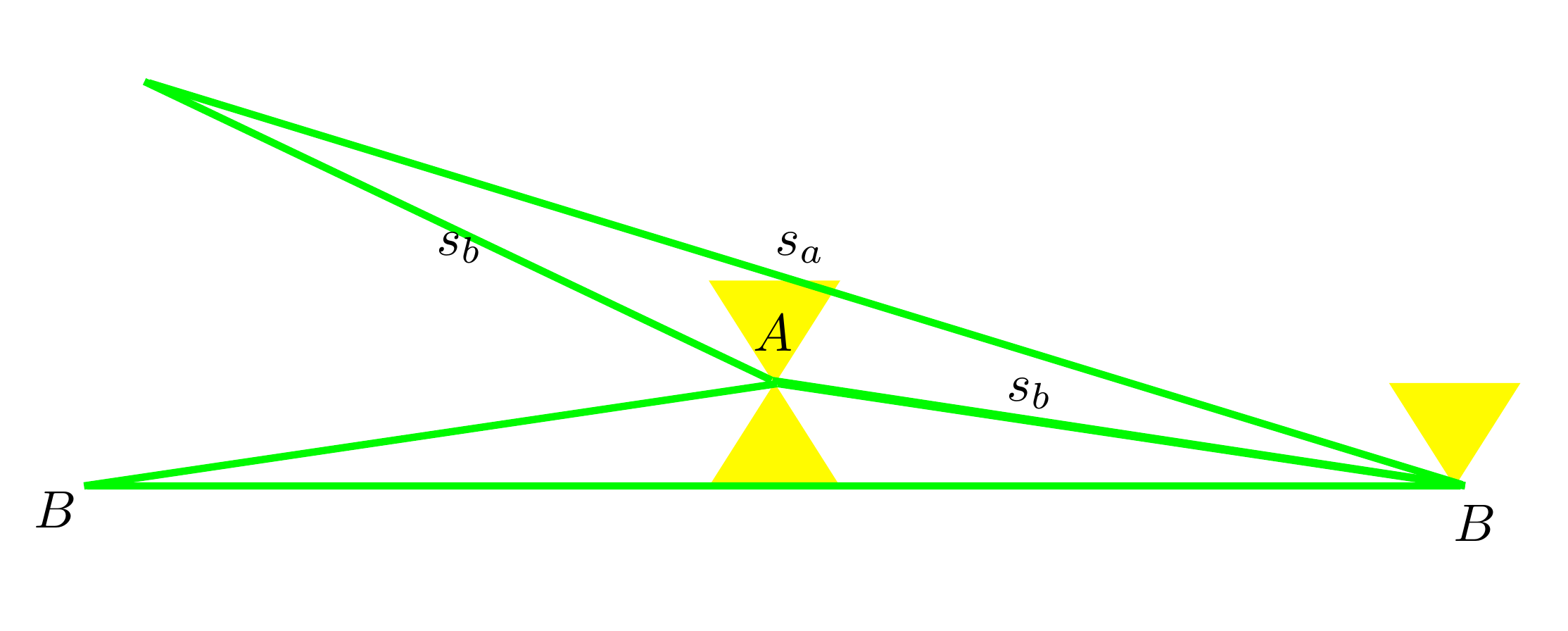}
        \caption{}
        \label{fig:spacelike_pair}
    \end{subfigure}
    \hfill
    \begin{subfigure}[c]{1\columnwidth}
        \centering
        \includegraphics[width=\linewidth]{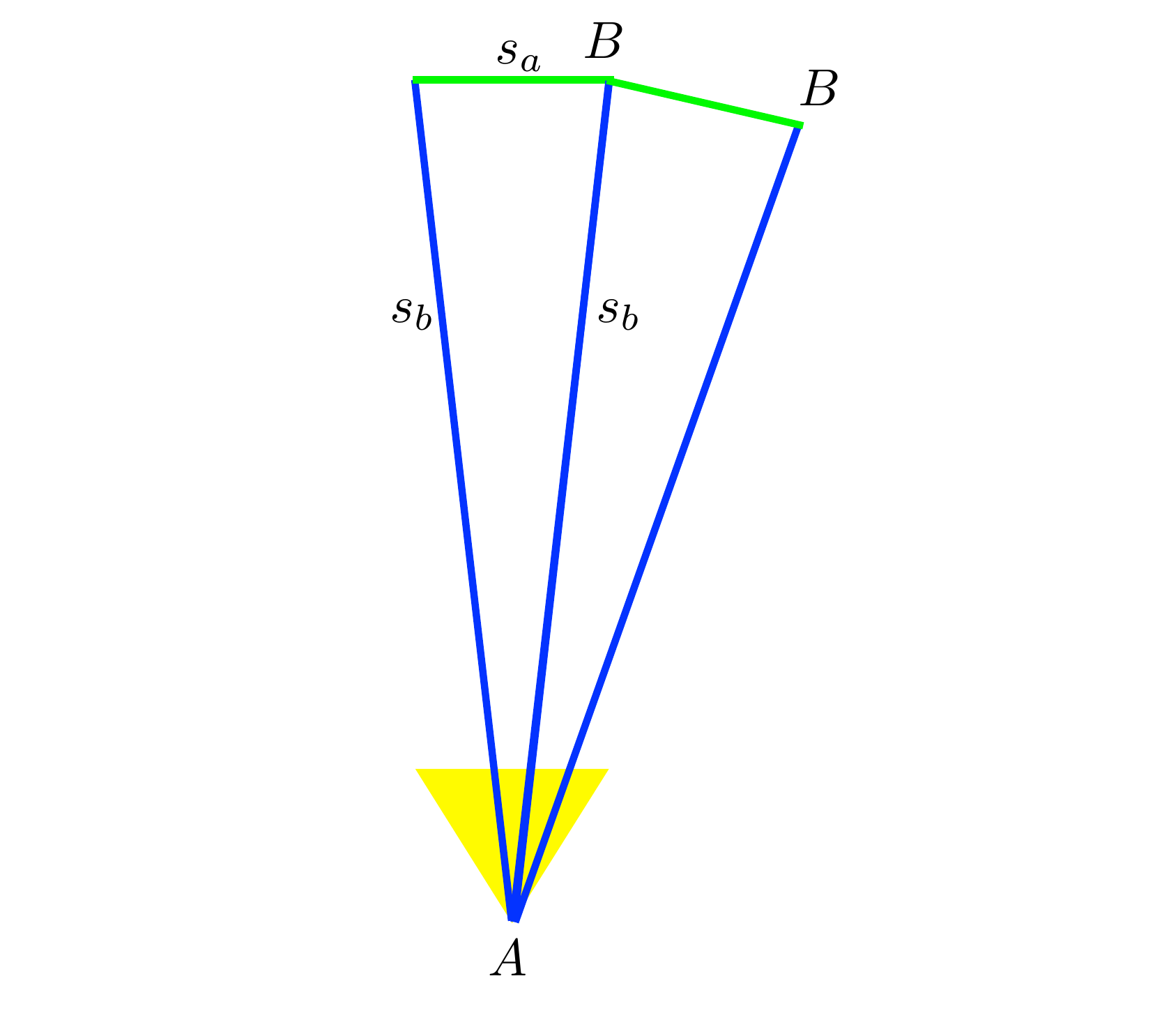}
        \caption{}
        \label{fig:timelike_pair}   
    \end{subfigure}
    \caption{Embedding of two triangles with squared edge lengths $s_a>0,s_b,s_b$ into Minkowski space. In the case (a) the $s_b$-edges are spacelike. If the light cone at vertex $A$ in the lower triangle is a past one, the light cone at vertex $A$ in the upper triangle has to be a future one and vice versa. In the case (b) the $s_b$ edges are timelike. Here the light cone at vertex $A$ does cover both triangle tips and therefore is on both triangles either future or past directed.}
\end{figure}

We will illustrate each of these points with examples:

~\\
(a): Consider two neighboring triangles sharing an $s_b$-edge in the first triangulation of the 2-sphere we considered in Section \ref{Sec:LargeBulk}. The triangles have the same geometry, given by one spacelike edge of squared length $s_a>0$, and two edges of length $s_b$. We glue these two triangles along an $s_b$-edge, so that vertices of the same type are identified with each other. FIG. \ref{fig:spacelike_pair}  illustrates the gluing and the embedding of the pair of triangles into Minkowski space, in the case that the $s_b$-edges are spacelike (FIG. \ref{fig:spacelike_pair}) and in the case that the $s_b$-edges are timelike (FIG. \ref{fig:timelike_pair}). Consider the light cone(s) at the $A$-vertex: in case that $0<s_b<s_a/4$, each triangle contains at the $A$-tip one light cone. If the light cone in the first triangle is a past cone, the light cone in the other triangle has to be a future cone, and vice versa. In the case $s_b<0$, the angles at the $A$ vertex include only timelike directions. We have a partial light cone at the $A$-vertex, which is either a future or a past light cone. The time flows in the two triangles are in the same direction. 

\begin{figure}[h!]
    \centering
    \begin{subfigure}[c]{1\columnwidth}
        \centering
        \includegraphics[width=\linewidth]{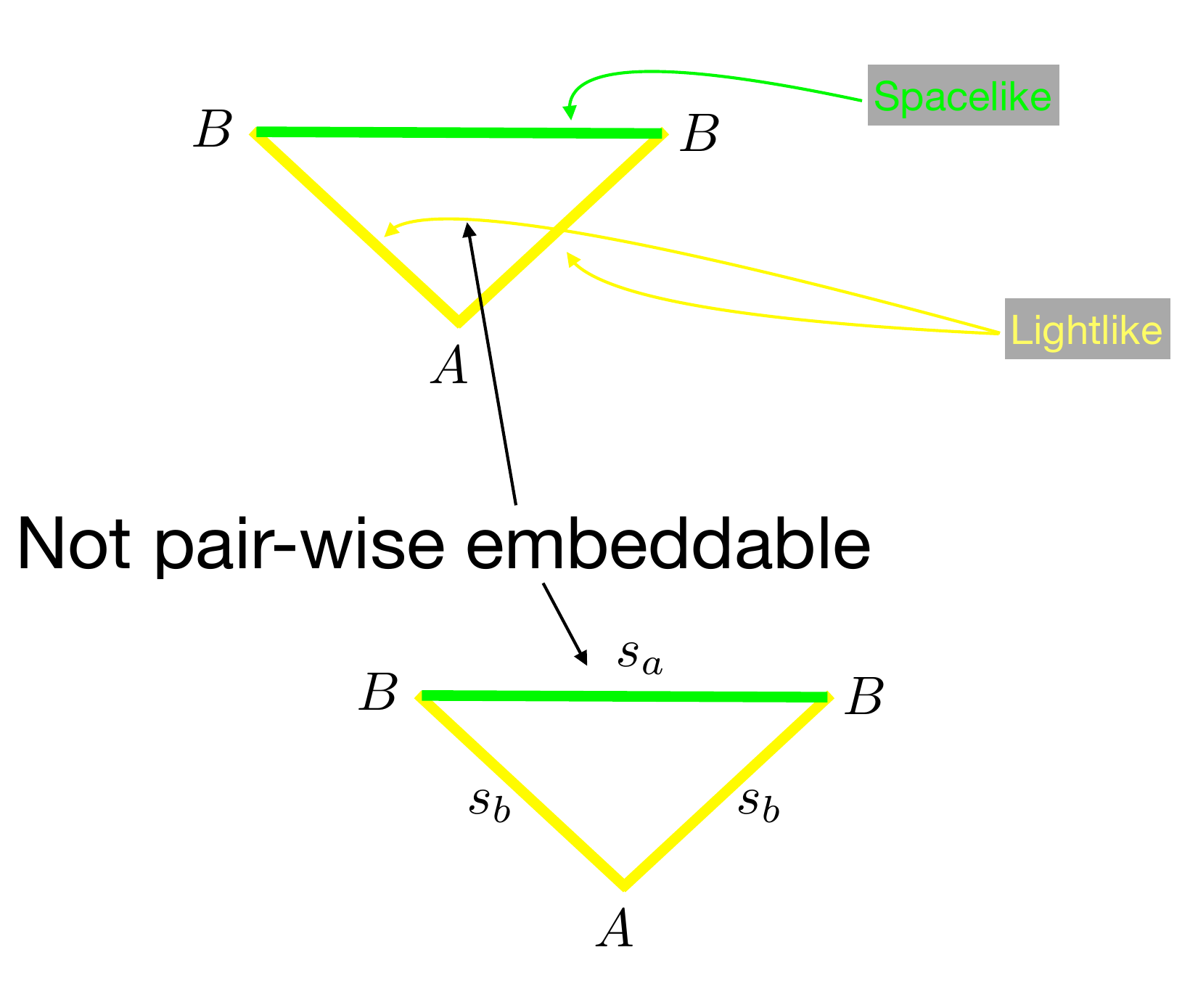}
        \caption{Two triangles which cannot be glued along the null edge $(AB)$, if we insist on positive relative spacetime orientation. We have chose the lower triangle's $AB$ edge to \emph{look} smaller purposefully.}
        \label{fig:null_pair}
    \end{subfigure}
    \hfill
    \begin{subfigure}[c]{1\columnwidth}
        \centering
        \includegraphics[width=\linewidth]{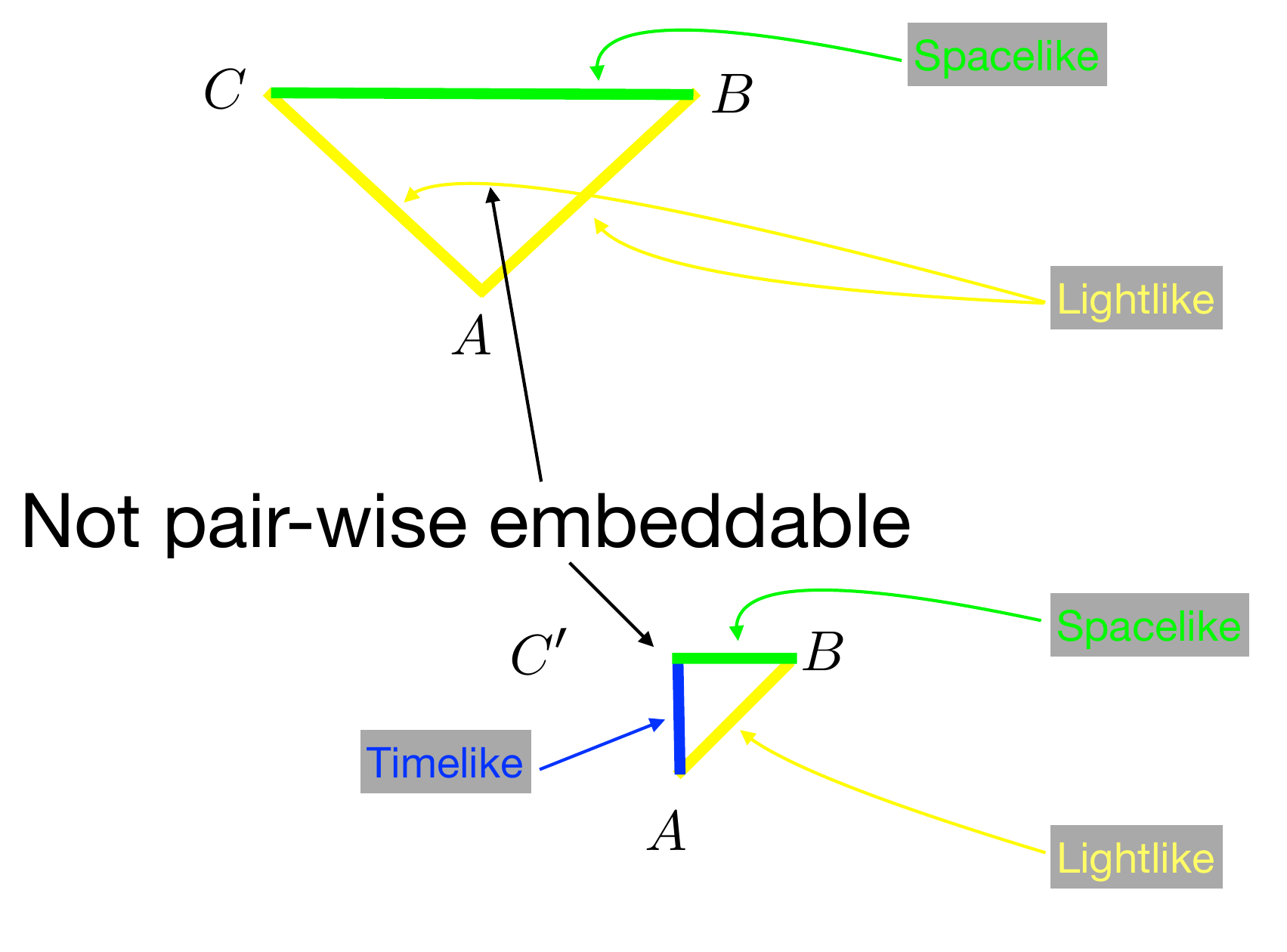}
        \caption{Less symmetric example of a pair of triangles that is not pairwise embeddable. We have chosen the lower triangle's $(AB)$ edge to \emph{look} smaller purposefully.}
        \label{fig:nonpairwise_asymmetric}   
    \end{subfigure}
    \caption{Examples of triangles glued along a null edge.}
\end{figure}

What happens in the case that the $s_b$-edges are null? The previous discussion shows that the transition from the spacelike to the timelike case has to be discontinuous. Indeed, this is an example where we \emph{cannot} embed the two neighboring triangles into Minkowski space, with both triangles either positively or negatively oriented. See \ref{fig:null_pair}.

Another example of triangles are not pairwise embeddable is shown in FIG. \ref{fig:nonpairwise_asymmetric}.

~\\
(b): Examples of triangulations which cannot be equipped with a pairwise embeddable chrono-topology come again from the first type of triangulations of the 2-sphere, considered in Section \ref{Sec:LargeBulk}. In the case that the $s_b$-edges are spacelike, we discussed above that the two light cones at a $B$-vertex of two neighboring triangles are future directed and past directed, respectively. That is, if one goes around the $B$-vertex, the light cones are alternating between being future and past directed. But this is only possible if we have an even number $N$ of light cones. If $N$ is odd, we cannot construct a pairwise embeddable chrono-topology. 

Similarly, the first type of triangulations of the 3-sphere we constructed has, in the case of spacelike $s_b$-edges, either an even or odd number of tetrahedra shared by a given $s_b$-edge. Here we have as necessary condition for a pairwise embeddable chrono-topology, that this number is even. This excludes the triangulations which are based on the 2-sphere triangulation given by the boundary of the tetrahedron and the boundary of the icosahedron.

~\\
(c): This point can again be illustrated by the first type of triangulations of the 2-sphere, considered in Section \ref{Sec:LargeBulk}.  In the case that the $s_b$-edges are spacelike, the time-orientations of two triangles sharing an $s_b$-edge are opposite to each other. The $B$-vertices are singular of yarmulke type. More precisely we have closed timelike curves going around these vertices in arbitrary small neighborhoods around these vertices. As we previously said, the number of triangles around a $B$-vertex needs to be even. 

In the case that the $s_b$ edges are timelike, two triangles sharing an $s_b$-edge have the same time orientation. There are only timelike direction emanating from a $A$-vertex, and at a given vertex these are either all future or all past directed. Now there is also a pairwise embeddable chrono-topology if we have an odd number of triangles around a given $B$-vertex.

Note the global change in time flow going from the first to the second case: For spacelike $s_b$ time is periodic and we have closed timelike curves, for timelike $s_b$ we have a non-periodic time flow `starting' at one $A$-vertex and ending at the other $A$-vertex. This illustrates why the name chrono-topology was chosen.

~\\

Another example of triangulations without a pairwise embeddable chrono-topology is provided by the second type of triangulations of the sphere, considered in Section \ref{Sec:LargeBulk}. These arose as a boundary of a 3D triangulation. In the case that the initial tetrahedron in this construction had Lorentzian signature, this 3D triangulation can be embedded into 3D Minkowski space. Remember that all triangles in the boundary of this 3D triangulation are timelike. With the embedding one induces a time orientation for each of these triangles. These time orientations however, do not lead to a pairwise embeddable chrono-topology: The 2D triangulation includes spacelike edges, whose inner points carry two light cones, which are either both future directed or both past directed: in the embedding 3D Minkowski space the two (timelike) triangles sharing such a spacelike edge cut through the same light cone. 
Trying to choose pairwise embeddable time orientations, we see that we cannot define a consistent global assignment, that is 
a pairwise embeddable chrono-topology does not exist for these 2D triangulations.

\begin{figure*}
    \centering
    \includegraphics[width=1\textwidth]{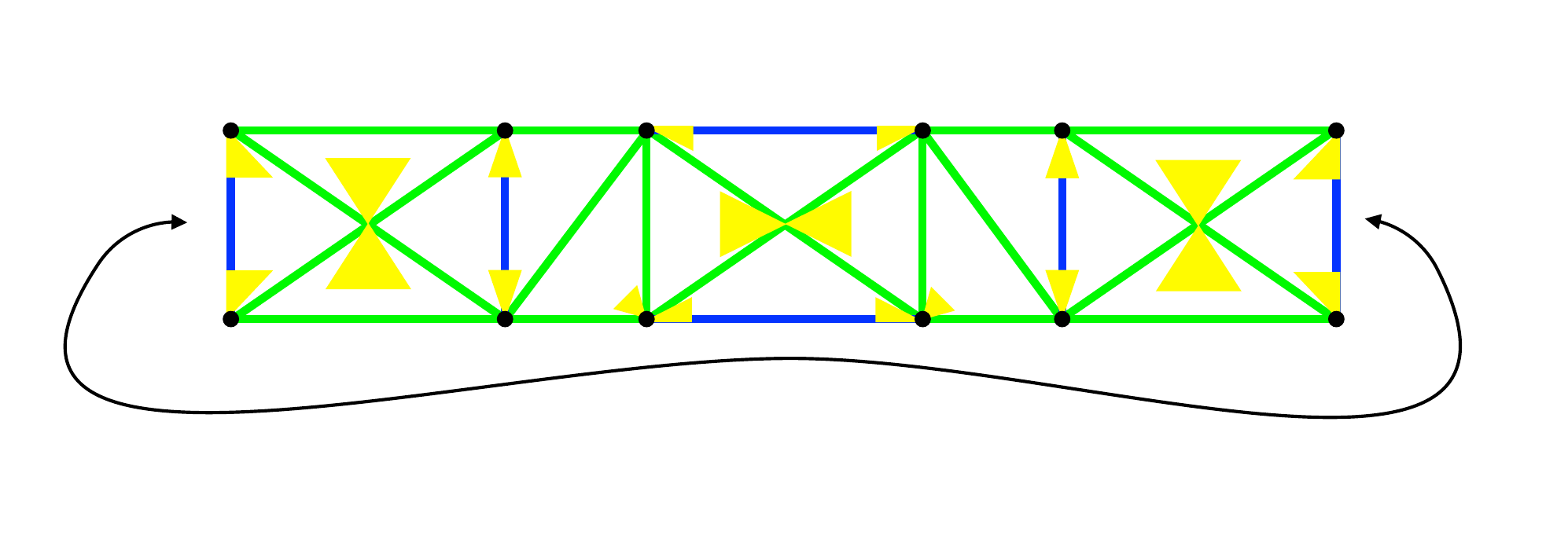}
    \caption{Example of a (portion of a) geometrized triangulation that is not time orientable —in the standard sense of \cite{Wald:1984rg}.}
    \label{fig:non_orientable_spacetime}
\end{figure*}

A further example (which has a well known continuum analogue) with a global obstruction for a pairwise embeddable chrono-topology, can be found in FIG. \ref{fig:non_orientable_spacetime}.

These examples illustrate that the conditions for a pairwise embeddable chrono-topology include both local and global aspects.

~\\

Above we noted that pairwise embeddability might be violated only if the two $d$-simplices share a null face. We will now provide a proof of this statement.  Given a spacelike or timelike $(d-1)$-dimensional face, we can embed this face into a spacelike or, respectively, timelike hyperplane of $d$-dimensional Minkowski space. Given one of the $d$-simplices we embed its remaining vertex into the $d$-dimensional Minkowski space, such that we reproduce the squared edge lengths of the remaining edges. For a given set of edge lengths (of a realizable non-degenerate simplex) there are two such embeddings, that is two sets of vertex coordinates. Once one has found one (which exists as we assume that the simplex is realizable), the other embedding can be found by applying a reflection with respect to the spacelike or timelike hyperplane. This reflection changes the spacetime orientation of the simplex. Applying this reasoning to both $d$-simplices we see that they can be embedded into Minkowski space with all (four) possible choices of spacetime orientations. Note that for the definition of pairwise embeddability above we only allowed a positive relative orientation. 

In order to understand the case with a shared null face, let us restrict ourselves to $2$-simplices, that is, triangles. Firstly we note that a reflection at a null-hypersurface cannot be defined as an isometry, as the normal to a null-hypersurface lies inside the null-hypersurface.\footnote{If we apply a reflection with respect to the $x=t$ line as we would in the Euclidean plane, it would change timelike directions into spacelike ones and vice versa and this reflection would not define an isometry.}  Thus, if there is an embedding of two triangles sharing a null edge, we cannot change its relative orientation by reflection along the null edge. 

Indeed, the conditions that two triangles with a shared null edge can be embedded with a fixed relative orientation include more than the demand that the geometry of the shared edge agrees. To explain this consider the two triangles after gluing. Pick one of the two vertices of the shared null edge. If we fix the relative orientation to be positive, the null edge has to separate spacelike directions from timelike directions emanating from this vertex, see FIG. \ref{fig:pairwise_embeddability2} (b).

Thus two realizable triangles sharing a null edge can be embedded into the Minkowski plane, but only with one choice of relative spacetime orientation. 

This discussion can be generalized to higher dimensional simplices by projecting the two simplices onto a plane. This plane is the one spanned by the two independent vectors, which are not orthogonal to the null direction in the shared null face.  We are then left with two two-dimensional triangles sharing a null edge, and can apply the reasoning we made above.

\subsection{Causality data \label{ssec:causality_data}}

Here we are going to discuss a way to encode the causality relations within a Lorentzian triangulation, and some of the constraints these causality data should possibly satisfy. 

We begin by noting that a given Lorentzian $d$-dimensional triangulation without null or degenerate simplices we can use the Cayley-Menger determinant of each $k$-simplex with $0< k\leq d$ to determine its spacelike/timelike nature. $0$-simplices, that is vertices, are by definition spacelike and all $d$-simplices are timelike. Note that a spacelike simplex cannot have timelike subsimplices. 

We might also encode future/past relationships by associate to each pair of a $k$-dimensional spacelike simplex $\sigma$ contained in a $(k+1)$-dimensional timelike simplex $\tau$ a sign $\epsilon$, which is $\epsilon=+1$ if $\sigma$ is a future boundary of $\tau$  and $\epsilon=-1$ if $\sigma$ is a past boundary of $\tau$. For abstract triangulations, these $\epsilon$'s might be thought of as additional degrees of freedom we add to spacelike (sub)simplices bounding timelike ones. When a simplex is embedded $\epsilon=\pm 1$  can be understood as the outward pointing timelike normal to $\sigma$ being future/past directed. Thus, this degree of freedom resolves an ambiguity we have when embedding a simplex just with Regge data.

The future past relationship have a natural set of constraints that follow from the causal data follow from the fact that the triangle inequalities guarantee, that we can embed a given $d$-simplex into $d$-dimensional Minkowski space:
\begin{itemize}
\item[(a)]
 If a timelike $(k+1)$-simplex $\tau$ has only spacelike $k$-simplices $\sigma_i$, with $i=1,\ldots,(k+2)$ then there is at least one $k$-simplex with $\epsilon(\sigma_i\subset \tau)=+1$ and at least one with $\epsilon(\sigma_i\subset \tau)=-1$. This follows from the fact that the normals to the $\sigma_i$ (weighted with their volume) are timelike and have to sum to zero. 
 \item[(b)]
 Consider a spacelike $k$-simplex $\sigma$ shared by a spacelike $(k+1)$-simplex $\sigma'$ and a timelike $(k+1)$-simplex $\tau$, which are all subsimplices of a timelike $(k+2)$-simplex $\tau'$, that is $\sigma \subset \sigma', \tau \subset \tau'$. Then $\epsilon(\sigma\subset \tau)=\epsilon(\sigma'\subset\tau')$. This just expresses that the time-orientation of a timelike $(k+2)$-simplex is imposed onto the timelike $k$-simplices it contains. 
 \item[(c)] The timelike edges of any $k$-simplex must form a comparability graph, that is the graph given by the timelike edges must be transitively orientable.\footnote{All graphs with four or fewer vertices can be transitively oriented. Therefore the rule starts to restrict the type of edges starting with 4-simplices. \emph{e.g.} the 5-vertex cyclic graph cannot be a comparability graph.}  Indeed, for a given timelike edge $e$ with two vertices $v,v'$ we have (because of (a)) that $\epsilon(v\subset e)=-\epsilon(v'\subset e)$. We can thus equip the edges with a  time orientation and the vertices of the $k$-simplex with a partial time order. This partial order has to be acyclic and transitive. That is, if $v,v'$ and $v''$ are vertices of the timelike edges in the $k$-simplex and we have $v\prec v'$ and $v'\prec v''$, the edge between $v$ and $v''$ has to be timelike and appropriately time-oriented.
\end{itemize}
We note that these constraints are not sufficient to guarantee the existence of a Lorentzian $d$-simplex.  

Consider a pair of neighboring $d$-simplices. If we demand  a positive relative spacetime orientation, then we have as necessary condition that
\begin{itemize}
\item[(d)]
 for any spacelike $(d-1)$-simplex $\sigma$, shared by two (necessarily timelike) $d$-simplices $\tau$ and $\tau'$ we have $\epsilon(\sigma\subset\tau)=-\epsilon(\sigma\subset\tau')$.
\end{itemize}
By contrast, if we have a pair of adjacent $d$-simplices with negative relative spacetime orientation, which share a spacelike $(d-1)$-simplex, we have $\epsilon(\sigma\subset\tau)=\epsilon(\sigma\subset\tau')$. The spacetime orientation data can be encoded into a sign associated to each $d$-simplex.

From the causality data we can reconstruct the number of light ray crossings included in a dihedral angle at a spacelike bone $\sigma_b$ between two  $(d-1)$-simplices $\rho$ and $\rho'$ in a $d$-simplex $\tau$.  To determine the dihedral angle one projects $\tau$ onto the plane orthogonal to $\sigma_b$. This results in a timelike triangle, in the following referred to as wedge, with the projected $\sigma_b$ resulting in a vertex, and $\rho$ and $\rho'$ projected to adjacent edges. Note  that these edges inherit the spacelike/timelike type from $\rho$ and $\rho'$, respectively. 

There are three combinations of spacelike/timelike types for $\rho$ and $\rho'$ in $\tau$:
\begin{itemize}
\item  Both $\rho$ and $\rho'$ are spacelike. If $\epsilon(\rho\subset \tau)=-\epsilon(\rho'\subset \tau)$, we have a so-called thin wedge and the dihedral angle does not contain a light ray crossing. If $\epsilon(\rho\subset \tau)=\epsilon(\rho'\subset \tau)$ we have a thick wedge and the dihedral angle contains two light ray crossings, and more precisely,  either a past (if $\epsilon(\rho\subset \tau)=+1$) or future light cone (if $\epsilon(\rho\subset \tau)=-1$). 
\item Both $\rho$ and $\rho'$ are timelike. If $\epsilon(\sigma_b\subset\rho)=+1$ then the edge of the wedge corresponding to $\rho$ (pointing away from $\sigma_b$) is past-directed, and if $\epsilon(\sigma_b\subset\rho)=-1$ the edge is future-directed. Thus if $\epsilon(\sigma_b\subset\rho)=\epsilon(\sigma_b\subset\rho)$ we have a thin (future or past-directed, depending on the sign) wedge without light ray crossings, and if $\epsilon(\sigma_b\subset\rho)=-\epsilon(\sigma_b\subset\rho)$ we have a thick wedge with two light ray crossings, with one future and one past directed light ray. 
\item One of the $(d-1)$-simplices is spacelike and the other is timelike. Such wedges include one light ray crossing. The light ray is future/past directed if the timelike edge of the wedge is future/past directed.
\end{itemize}

Let us restrict to Lorentzian triangulations where all pairs of $d$-simplices have a positive relative spacetime orientation.\footnote{One can adjust to the more general case if one takes into account that the (relative) spacetime orientation of the $d$-simplices is imposed onto the wedges which result from their projection. \emph{e.g.} dihedral angles from positive and negative oriented wedges enter with opposing sign into the deficit angle.} In this case we notice that we can hinge on the points above to check whether a bone is  light-conical singular or not by counting the number of light cones directly from the causality data. In fact the causality data gives a finer notion of regular light cone structure, as it can check if these four light rays determine one future and one past light cone.

This might suggest that the causality data may provide a very efficient way to determine when a configuration has a light-conical singularity. However, observe that the geometry puts constraints on the causality data (and vice versa): For example, if we are in a situation like the one described in the first point, geometry also determines whether the wedge is thick or thin. So there are compatibility conditions between the Regge geometric data and the causality data.

It might be possible to also formulate edge and vertex causality (see Section \ref{Sec:Other}) as conditions on the causality data. We leave this question for future research.

We note that the causality data proposed here coincide  with the data proposed in the spin foam literature \cite{Markopoulou:1997wi, Livine:2002rh, Bianchi:2021ric} if one restricts to the case considered in the four-dimensional spin foam literature so-far, namely that $(i)$ all $(d-1)$-simplices are spacelike and $(ii)$ all pairs of $d$-simplices have positive relative spacetime orientation. In this case $\epsilon(\sigma\subset\tau)=-\epsilon(\sigma\subset\tau')$ for a $(d-1)$-simplex $\sigma$ shared by the $d$-simplices $\tau$ and $\tau'$. We can therefore replace the $\epsilon$ with an orientation for the edges of the dual graph —whose vertices represent the $d$-simplices and edges the $(d-1)$-simplices. It is noted in \cite{Markopoulou:1997wi, Livine:2002rh, Bianchi:2021ric} that the dual graph obtains in this way the structure of a causal set (if there are no cycles), but there are so far no further discussions of regularity conditions. Above we discussed a condition which ensures that there are no light-conical singularities. For the dual complex with oriented edges it amounts to a condition based on the two-dimensional cells of the dual complex. Such cells are bounded by a cycle of oriented edges, and the condition is that this cycle  includes exactly two orientation changes.

Similarly, one can check that this spin foam-like causality data ensure there is an even number of light cones at a bone.\footnote{We are grateful to Wojciech Kaminski for pointing this out.}

Finally, let us emphasize that we restricted part of our (explicit) discussion to positive relative spacetime orientation.  However, a sum over both spacetime orientations for each $d$-simplex seems to be necessary in order to obtain a path integral which acts as a projector onto the Hamiltonian and diffeomorphism constraints \cite{Halliwell:1990qr,Noui:2004iy}.\footnote{We expect that to obtain a projector for the four-dimensional case one needs in general a continuum-limit \cite{Bahr:2009qc,Asante:2022dnj}. There might be exceptions, \emph{e.g.} if one considers so-called tent-moves for four-valent vertices \cite{Dittrich:2009fb,Dittrich:2011ke}.} It is also not clear whether a restriction to a given spacelike/timelike character for $d$-simplices or for the edges results in a path integral, which acts as a projector onto the Hamiltonian and diffeomorphism constraints.

\section{Discussion \label{sec:discussion}}

The construction of a Lorentzian path integral over geometries requires the specification of which kind of geometries to integrate over. A particularly important choice, which can heavily influence the outcome of the path integral is, which kind of light cone and causal structures are allowed. Previous work indicates that allowing so-called light-conical singularities, that is irregularities in the light cone structure with codimension-two support, is crucial in order to derive thermodynamic quantities from the Lorentzian path integral \cite{Colin-Ellerin:2021jev,Marolf:2022ybi,Dittrich:2024awu}.\footnote{\cite{Asante:2021phx} shows that including light-conical singularities can be also crucial in Regge models for quantum cosmology. \cite{Asante:2021phx} discusses a path integral set-up for the no-boundary wavefunction and finds that in order to obtain results matching the continuum one needs to also integrate over configurations with light-conical singularities. The related branch cuts in the Regge action correspond to an essential singularity for the effective action (after integrating out the spatial degrees of freedom) in the continuum.} One important reason for this is that light-conical singularities imply imaginary contributions for the Lorentzian action, which are also accompanied by branch cuts. The sign for these imaginary contributions depends on a choice of integration contour around these branch cuts \cite{Asante:2021phx}, the contour therefore determines whether these configurations are suppressed or enhanced in the path integral.

Assuming we do allow light-conical singularities in the path integral (and provide a description for the choice of integration contour), can we expect to obtain reasonable results? The answer heavily depends on (a) how frequently such light-conical singularities appear and (b) whether they can appear with an unbounded integration domain. In this work we explored these questions. 

Random sampling of the four-dimensional Regge configuration space in different set-ups shows strong evidence that configurations with light-conical singularities make up the majority of the configurations. In particular if we consider bones which are shared by five or more 4-simplices, light-conical singularities of trouser type appear much more often than yarmulke like singularities. Counting the number of light rays included in Lorentzian deficit angles attached to bones, we found in particular, that the maximum of the distribution occurs when the number of light rays matches the number of 4-simplices shared by the bone in question. Most triangulations feature bones with this number being different from four, as illustrated by our example of the triangulation of the hypercube. Restricting all tetrahedra in the four-dimensional triangulation to be spacelike (as done for the so-called EPRL/FK spin foam models \cite{Engle:2007wy,Freidel:2007py} in their original formulation ) can significantly influence the statistics of light-conical singularities, and shift the maximum of the distribution to yarmulke-like singularities.   

Configurations with light-conical singularities with an unbounded integration domain can be also found easily. We note that the spine and spike configurations considered here arise from refinement operations \cite{Borissova:2024pfq,Borissova:2024txs} and in particular from Pachner moves \cite{PACHNER1991129}. For three-dimensional triangulations we found only light-conical singularities with actions where the yarmulke contributions would dominate.\footnote{This does not exclude that trouser-like singularities with unbounded integration domain exist. But we provided a complete listing for spine and spike configurations.} For four-dimensional triangulations spine configurations do not lead to light-conical singularities —at the triangles shared by the `spine'. But we found spike configurations which can include either yarmulke or trouser type singularities, or both types or no light-conical singularities.

A main contribution of this work is an algorithm that produces Regge geometries that satisfy the triangle inequalities. This was a main tool we used to explore the Regge path integral configuration space. We believe that this algorithm has many more applications beyond this work. 

We furthermore discussed to which extent triangulations equipped with realizable geometrization do or do not admit a consistent causal structures. We identified a novel gluing rule for the case of codimension-one simplices which are null. The two simplices which share such a null simplex can be only glued together either with positive or with negative relative orientation. (In contrast, two $d$-simplices sharing either a timelike or spacelike $(d-1)$-simplex can be glued together with both relative orientations.)

We introduced the notion of pairwise embeddable chrono-topologies and saw that one can easily produce Lorentzian triangulations which do not admit such chrono-topologies. Moreoever we saw that varying the geometric data on a triangulation with fixed connectivity can change its chrono-topology, or change whether there exist a  pairwise embeddable chrono-topology. We presented a way to equip a triangulation with causality data and discussed various constraints that either arise because the $d$-simplices are Lorentzian, or one has a pairwise embeddable chrono-topology, or one forbids light-conical singularities.

The results we found show that for a general triangulation we have to expect an abundance of light-conical singularities and more generally causal irregularities. Whereas trousers seem to be, in general,  entropically preferred, yarmulkes (and, at least in four dimensions, also trousers) can appear with an unbounded integration domain in the context of simple refinement operations. An open question for future research is whether such configurations, even when exponentially enhanced, can be canceled out by destructive interference, as observed in \cite{Dittrich:2024awu} for a specific example. 

Another possibility to suppress irregularities in the causal structure is to introduce more constraints for the geometric data to be integrated over.

There are only a few proposals for such constraints. \cite{Tate:2011ct}  suggested to imitate the structure of Causal Dynamical Triangulations for the Regge path integral: that is imposing a slicing of the triangulation into spacelike hypersurfaces (which include only spacelike building blocks), with these spacelike hypersurfaces connected by only timelike edges.  This was hoped to suppress configurations maximizing the conformal factor fluctuations (as could then Wick rotate and apply Monte Carlo techniques), see \cite{Ito:2022ycc,Ito:2025yop} for further explorations of this question. This proposal has the advantage to have a simple restriction rule, but the possible disadvantage of having a cut-off in configuration space, which does not need to coincide with the (dis)appearance of any kind of causal irregularities.\footnote{Consider for instance a regular two-dimensional lattice made of rectangles with timelike side lengths $|t|$ and spacelike side lengths $|s|$. Subdivide each of the rectangles into two triangles by introducing diagonals with length square $d^2$. The restriction rule would only allow spacelike diagonals, although configurations with $|t|<|s|$ (for which the diagonals are spacelike) are perfectly reasonable.} An alternative would be to disallow irregularities in the light cone structure. Disallowing light-conical singularities might be quite straightforward within the random sampling algorithm employed in this work. This algorithm proceeds by exploring the circuits of one bone after the other. One can therefore keep track of the number of light rays around a given bone and discard configurations which have more or less than four light rays. 

The causality data we introduced allow also to determine the number of light rays (and more specifically whether they are future or past directed) around a bone, which can be used to discard configurations with light-conical singularities. 

It is generally expected that in order for the path integral to project onto configurations satisfying the Hamiltonian and diffeomorphism constraints (possibly after having taking a continuum limit \cite{Dittrich:2014ala, Asante:2022dnj}), one needs to sum over positive and negative spacetime orientations for the $d$-simplices.  Likewise it is not clear whether restricting to a given chrono-topology or to a given prescription of spacelike/timelike character for the subsimplices will lead to a path integral acting as projector onto the constraints. This question could be investigated starting with local time-evolution moves, \emph{e.g.} so-called tent- or Pachner-moves \cite{Barrett:1994ks, Dittrich:2011ke}, which can be imposed via a path integral. 

Another possibility for suppressing  causally irregular configurations is to include matter couplings. Matter may exhibit a singular behavior on such configurations. A related argument appears to the choice of contour around the branch cuts which arise for light-conical singularities. This forces a complexification of the metric, which can either act as a regulator for matter fields or enhance divergencies. This leads one to argue for a suppression of trousers and an enhancement of yarmulkes, see \cite{Halliwell:1989dy, Dowker:1997hj, Kontsevich:2021dmb,Witten:2021nzp}. If adopted, it would be helpful to clarify the fate of yarmulke light-conical singularities with unbounded integration support. As discussed here and in \cite{Borissova:2024pfq,Borissova:2024txs} these appear for Pachner move configurations, which are quite accessible for a numerical analysis.

\appendix{

\section{Jacobi determinants}\label{App:Jac}

Let us first consider the Jacobi determinant for the dihedral construction. Given a $d$-dimensional simplex $\sigma^d$ (with $d>2)$ and a two-dimensional bone in this simplex we compare the variation of the squared length $s$ of the edge opposite the bone $b$ to the variation of the squared length $s_{\rm PT}$ of an edge $e_{\rm PT}$ in the `projected' triangle, which results from projecting out $b$ from $\sigma^d$. The edge $e_{\rm PT}$ is opposite the vertex $v_b$ which results from projecting the bone $b$ in $\sigma^d$. By definition the dihedral angle $\theta_{v_b}$ at this vertex in the projected triangle is equal to the dihedral angle $\theta_b$ at the bone $b$ in $\sigma^d$. 

We consider the case that $\sigma^d$ is Lorentzian and the bone $b$ is Euclidean, and the dihedral angle is therefore Lorentzian. All other cases can be considered similarly. 

We will employ the derivative of the dihedral angle $\theta_h$ at a bone (or hinge) $h$ with respect to the squared length $s_{\bar{h}}$ of the edge opposite the hinge. For an $n$-dimensional simplex (and for Lorentzian angles), this is given by (see \cite{Dittrich:2007wm} for Euclidean simplices, and \cite{Borissova:2023izx} for Lorentzian simplices)
\ba\label{Dertheta}
\frac{\partial \theta_h}{\partial s_{\bar{h}} } &=& -\imath \frac{1}{2(n-1)n}\frac{\sqrt{ \mathbb{V}_{\bar h} }}{\sqrt{\mathbb{V}_n}} \q .
\ea
Here, $\mathbb{V_{\bar h}}$ is the signed squared volume of the hinge, $\mathbb{V}_n$ is the signed squared volume of the $n$-simplex $\sigma^n$. 

On the one hand we apply (\ref{Dertheta}) to determine the variation of the dihedral angle $\theta_b$ in the $d$-simplex $\sigma^d$ with respect to the squared edge length $s$. On the other hand we apply (\ref{Dertheta}) to determine the variation of $\theta_{v_b}$ in the projected triangle with respect to $s_{\rm PT}$. Using $\theta_b=\theta_{v_b}$ and that $(d!)^2\mathbb{V}_d=((d-2)!)^2 \mathbb{V}_b \times 4 \mathbb{V}_{\rm PT}$ one finds 
\ba
d s&=& ds_{\rm PT} \q ,
\ea
that is a trivial Jacobi determinant. 

The other relation we want to consider is the following: We have given $d$ vertices of a $d$-simplex in $d$-dimensional flat space, and want to add another vertex $v_0$  with coordinates $v_0^\alpha$. We can then express the measure in squared edge lengths of all the added edges in terms of the measure in the vertex coordinates (by computing the derivatives of $s_{0a}$ with respect to $v_0^\alpha)$ \cite{Baratin:2006yu, Dittrich:2011vz, Borissova:2023izx}:
\begin{align}
\sum_{\text{orient}} \prod^d_{a=1} d s_{0a}= 2^d d! \, V\, \prod^d_{\alpha=1} dv^\alpha_0 \q ,
\end{align}
where $V$ is the absolute volume of the $d$-simplex. This volume can be easily expressed as function of the coordinates or as function of the edge lengths.

$\sum_{\text{orient}}$ indicates a sum over the two orientations, which appears for the following reason: Having the new vertex to one side of the hyperplane defined by the remaining vertices defines one orientation, and having it on the other side leads to the opposite orientation. Excluding the degenerate case of the vertex situated inside the hyperplane (which is of measure zero), one assignment of edge lengths maps to two assignments of coordinates, one with positive and one with negative orientation.

}

\section*{Acknowledgements}

We are very thankful to Fay Dowker and Rafael Sorkin for questions and discussions on causal structures in the Regge framework, which were highly influential for Section \ref{sec:chrono}. JPA is supported by an NSERC grant awarded to BD.
Research at Perimeter Institute is supported in
part by the Government of Canada through the Department of Innovation, Science and Economic
Development Canada and by the Province of Ontario through the Ministry of Colleges and Universities.

\bibliographystyle{apsrev4-1}
\bibliography{references}

\end{document}